\documentclass[structabstract]{aa}  
\usepackage{graphicx}
\usepackage{txfonts}
\usepackage{natbib}
\bibpunct{(}{)}{;}{a}{}{,} 
\begin{document}
   \title{GMASS ultradeep spectroscopy of galaxies at $z \sim 2$}

   \subtitle{VII. Star formation, extinction, and gas outflows from UV spectra}

\author{M. Talia
\inst{1}
\and
M. Mignoli\inst{2} \and
A. Cimatti\inst{1} \and
J. Kurk\inst{3} \and
S. Berta\inst{3} \and
M. Bolzonella\inst{2} \and
P. Cassata\inst{5} \and
E. Daddi\inst{4} \and
M. Dickinson\inst{6} \and
A. Franceschini\inst{7} \and
C. Halliday\inst{8}
L. Pozzetti\inst{2} \and
A. Renzini\inst{9} \and
G. Rodighiero\inst{7} \and
P. Rosati\inst{10} \and
G. Zamorani\inst{2} 
}
          
\offprints{M. Talia\\
\email{margherita.talia2@unibo.it}}

\institute{Dipartimento di Astronomia, Universit\`a di Bologna,
Via Ranzani 1, I-40127, Bologna, Italy.
\and  
INAF- Osservatorio Astronomico di Bologna,
Via Ranzani 1, I-40127, Bologna, Italy
\and
Max Planck Institut f\"ur extraterrestrische Physik,
Postfach 1312, 85741 Garching bei M\"unchen, Germany
\and
CEA-Saclay, DSM/Irfu/Service d'Astrophysique, 91191
Gif-sur-Yvette Cedex, France
\and
Department of Astronomy, University of Massachusetts, 710 North Pleasant
Street, Amherst, MA 01003, USA
\and
NOAO-Tucson, 950 North Cherry Avenue, Tucson, AZ 85719, USA
\and
Universit\`a di Padova, Dipartimento di Astronomia,
vicolo dell'Osservatorio 2, I-35122, Padova, Italy
\and
INAF- Osservatorio Astrofisico di Arcetri, 
Largo E. Fermi 5, I-50125 Firenze, Italy
\and
INAF- Osservatorio Astronomico di Padova,
vicolo dell'Osservatorio 5, I-35122, Padova, Italy
\and
European Southern Observatory,
Karl-Schwarzschild-Strasse 2, D-85748, Garching bei M\"unchen, Germany
}

   \date{}

 
  \abstract
   {}
   {We use rest-frame UV spectroscopy to investigate the properties related to large-scale gas outflow, and to the dust extinction and star-formation rates of a sample of $z \sim 2$ star-forming galaxies from the Galaxy Mass Assembly ultradeep Spectroscopic Survey (GMASS).}
   {Dust extinction is estimated from the rest-frame UV continuum slope and used to obtain dust-corrected star-formation rates for the galaxies of the sample. A composite spectrum is created by averaging all the single spectra from our sample, and equivalent widths and centroids of the absorption lines associated with the interstellar medium are measured. Velocity offsets of these lines relative to the systemic velocity of the composite, obtained from photospheric stellar absorption lines and nebular emission lines, are then calculated. Finally, to investigate correlations between galaxy UV spectral characteristics and galaxy general properties, the sample is divided into two bins that are equally populated, according to each of the following galaxy properties: stellar mass, colour excess, star-formation rate, and morphology; then a composite spectrum for each group of galaxies is created, and both the velocity offsets and the equivalent widths of the interstellar absorption lines are measured.}
   {For the entire sample, a mean value of the continuum slope $<\beta>=-1.11 \pm 0.44$ (r.m.s.) was derived. For each galaxy, dust extinction was calculated from the UV spectrum and used to correct the flux measured at 1500 $\AA$ (rest-frame), and the corrected UV flux was then converted into a star-formation rate. We found our galaxies to have an average  $<SFR>~=52 \pm 48 ~M_\odot~ yr^{-1}$ (r.m.s.).
A positive correlation between SFR and stellar mass was found, in agreement with other works, the logarithmic slope of the relation being $1.10 \pm 0.10$.

Low-ionization absorption lines, associated with the interstellar medium, and measured on the composite spectrum, were found to be blueshifted, with respect to the rest frame of the system, which indicates that there is outflowing gas with typical velocities of the order of $\sim 100$ km/s. 
Finally, investigating correlations between galaxy UV spectral characteristics and galaxy general properties, we find a possible correlation between the equivalent width of the interstellar absorption lines and SFR, stellar mass, and colour excess similar to that reported to hold at different redshifts.}
   {}


   \maketitle
%
\section{Introduction}
\label{sec:Introduction}

In the past two decades, great effort has been devoted to the study of the Universe in the redshift range $1 < \emph{z} < 3$. Several works have shown that this is the epoch when a substantial fraction of galaxy mass assembly took place, and when there is a peak in the evolution of the star-formation rate density through cosmic time \citep{1996ApJ...460L...1L, 1996MNRAS.283.1388M, 2003ApJ...587...25D, 2006ApJ...651..142H, 2007ApJ...670..156D}. 
Observations appear to be consistent with a differential evolution of galaxies, according to which the most massive galaxies ($M > 10^{11} M_\odot$) formed their stars earlier and more rapidly than the less massive ones. This so-called \emph{downsizing} scenario, first introduced by \citet{1996AJ....112..839C}, was later confirmed in many works focused on the $M_\ast / L$ \citep{2004A&A...424...23F} and, especially, the specific star-formation rate at different redshifts \citep{2003ApJ...594L...9F, 2005ApJ...633L...9F, 2005ApJ...630...82P, 2006ApJ...640...92P, 2009A&A...504..751S, 2010A&A...518L..25R}. Another element has been added to this picture with the discovery of a population of distant ($1 < \emph{z} < 2$) massive early-type galaxies that have the spectral properties of passively evolving old stars with ages of $1 - 4$ Gyr \citep{2004Natur.430..184C}. The properties of these galaxies imply a passive-like evolution after a major, strong ($> 100~ M_\odot~ yr^{-1}$), and short-lived ($\tau \sim 0.1 - 0.3$ Gyr)\footnote[1]{$\tau$ indicates the scale time of star formation where its rate is $SFR \propto exp(-t/\tau)$.}, starburst episode occurred at $1.5 < \emph{z} < 3$.

A key role in the study of star-forming galaxies, in this critical redshift range, is played by the optical spectroscopic surveys that have been made possible by the advent of 8-10 m class telescopes. The observed optical regime samples the rest-frame UV range for galaxies at redshift $\emph{z} \sim 2$. The rest-frame UV spectrum contains much information about the star-formation activity of the galaxy, the hot young stellar population, and the physical, chemical, and dynamical state of the interstellar medium. These properties have been extensively studied in the local universe \citep[see][]{1993ApJS...86....5K, 1998ApJ...503..646H} and, most recently, higher redshift data have been collected from both large surveys \citep{2003ApJ...588...65S, 2004ApJ...602...51S, 2009ApJ...692..187W, steidel2010} and lensed single objects \citep{2000ApJ...528...96P, 2002ApJ...569..742P, quider2009, quider2010}.
The UV continuum of a star-forming galaxy (SFG) is dominated by radiation emitted by young massive O and B stars, therefore its UV luminosity could be used as a direct estimator of the ongoing star formation in a galaxy \citep{1998ARA&A..36..189K}. However, it is known that the rest-frame UV wavelength range is strongly affected by dust extinction. Although both the composition of dust grains and their distribution in galaxies are not yet completely understood, empirical extinction laws are typically used to remove the effects of dust extinction from the observed spectral energy distribution of star-forming galaxies \citep{2000ApJ...533..682C}. Furthermore, \citet{1994ApJ...429..582C} showed that the shape of the continuum can be fairly well approximated by a power law $F_\lambda \varpropto \lambda^\beta$, where $F_\lambda$ is the observed flux ($erg~s^{-1}~cm^{-2}~\AA^{-1}$) and $\beta$ is the continuum slope, and that there is a strong correlation between the slope and nebular extinction measured in the optical using the Balmer decrement. 
\citet{1999ApJ...521...64M} found a linear relation between the UV continuum slope, parametrized by $\beta$, and dust extinction at 1600 $\AA$, expressed in magnitudes. The relation, calibrated for a sample of local starburst galaxies, has also been found to hold at higher redshifts \citep{2006ApJ...644..792R, 2010ApJ...712.1070R}. 

Rest-frame UV spectra provide not only an estimate of the extinction-corrected star-formation rates, but also information about one of the principal mechanisms currently invoked by galaxy evolution models to quench the star formation in a galaxy, as well as enrich the inter-galactic medium (IGM) with metals: large-scale gas outflows. The UV spectrum of a star-forming galaxy, in the wavelength range between 1100 $\AA$ and 2000 $\AA$, is characterized by strong absorption lines, which originate from absorption of the stellar radiation by the interstellar medium (ISM) surrounding the regions of intense star-formation. These lines come from different elements, predominantly carbon, silicon, and iron, at different ionization stages. Some of them are the result of the combination of various mechanisms, such as stellar winds, in addition to the interstellar component, and therefore their profiles may be quite complicated to analyze. A common property of these lines is that they have been found to be blueshifted at velocities of about $\sim$ 100 km/s, with respect to galaxy rest-frame, in the spectra of galaxies in the local universe as well as at higher redshifts \citep{1998ApJ...503..646H, 2003ApJ...588...65S}, suggesting that the gas is flowing out from the galaxy. 
The equivalent width (EW) of these absorption lines has been found to be correlated with several galaxy properties \citep{1998ApJ...503..646H, 2003ApJ...588...65S, 2009ApJ...692..187W}; in particular, the lines are stronger in galaxies characterized by higher SFRs, higher masses, and higher dust extinctions, but it is not yet clear on which property the outflow parameters are most strongly dependent. These absorption lines have been shown to be saturated, therefore they cannot be used to infer chemical abundances (though, in some cases, lower limits to the column density may be estimated). \citet{1998ApJ...503..646H} points out that, in the case of a saturated line, the EW is only weakly dependent on the ionic column density, and more strongly dependent on the velocity dispersion of the absorbing gas. Therefore, the correlation between EW and dust extinction might imply that the velocity dispersion in the absorbing gas is larger in galaxies with more dust extinction. By analyzing the Na $\lambda\lambda 5890, 5896~\AA$ doublet in 32 far-IR selected galaxies, \citet{2000ApJS..129..493H} concluded that winds with larger velocity spreads are driven by galaxies with higher star-formation rates, which contain more dust. \citet{2003ApJ...588...65S} alternatively proposes that the correlation between the EW of interstellar absorption lines and dust extinction can be more directly explained by considering the covering fraction of dusty clouds.

Different models of galactic-scale winds in star-forming galaxies have been proposed over the years \citep{1985Natur.317...44C, 1999MNRAS.309..332T, 2000MNRAS.314..511S, 2009MNRAS.396L..90N, 2009astro2010S.289S, steidel2010}, but a general consensus remains to be found. It is generally accepted that galactic-scale winds originate from the disruption of bubbles of hot gas, created by a combination of the effects of supernovae and stellar winds in actively star-forming regions. Galactic winds are also believed to have a multi-phase gas structure, in principle providing observables at all wavelengths from X-ray to optical. The blue-shifted absorption features that can be found in the wavelength range between 1100 $\AA$ and 2000 $\AA$, originate in the warm phase of the outflow. The hot phase ($10^7 - 10^8$ K), emitting in the X-ray regime, is the "engine" of the wind and carries most of the energy. The exact role of the warm phase in the general structure of a galactic-scale outflow is still under debate. In some models, the expanding hot gas entrains fragments of the cooler ambient gas, leading to the formation of clouds moving at lower velocities than the main flow, in which the UV blue-shifted absorption lines originate. In other models, the UV lines originate in ambient gas pushed by the expanding bubble in the pre-wind phase. Eventually, it is not yet clear if the gas, though flowing out from the galaxy, is actually able to reach the IGM or if, instead, it is trapped in the extended haloes and then accreted back onto the galaxy. To discriminate between different models, much help should be provided by X-ray observations. Evidence of outflowing gas has often been detected in star-forming galaxies, but only in the UV and optical regimes (for a comprehensive review on galactic winds, see \citet{2005ARA&A..43..769V}). 

In this paper, we analyze the rest-frame UV spectra of a sample of 74 star-forming galaxies from GMASS (\emph{Galaxy Mass Assembly ultra-deep Spectroscopic Survey}), a survey expecially designed to search for and study both passive early-type galaxies and their star-forming progenitors at high redshift. We focus on determining the extinction-corrected star-formation rates from the UV luminosity, which is directly measured in the spectra. By using composite spectra, we also study the properties of the absorption spectral lines associated with the ISM, and investigate possible correlations with physical galaxy properties, such as stellar mass, SFR, and morphology. The paper is organized as follows: in Sect.\ref{sec:The sample}, we introduce the dataset and our selection criteria. In Sect.\ref{sec:Rest-frame UV spectroscopic features}, the principal spectral features of the spectra in our sample are described, with particular emphasis on the ones related to the search for large-scale outflows. In the following sections, we present our analysis of the sample, along with the results concerning the dust extinction inferred from the continuum slope, the SFR, and the UV absorption lines associated with the outflowing gas. Finally, we try to correlate the physical properties of the galaxies in our sample (such as star-formation rate and stellar mass) with UV line properties.

\section{The sample}
\label{sec:The sample}

\begin{figure}
\centering
\includegraphics[scale=0.4]{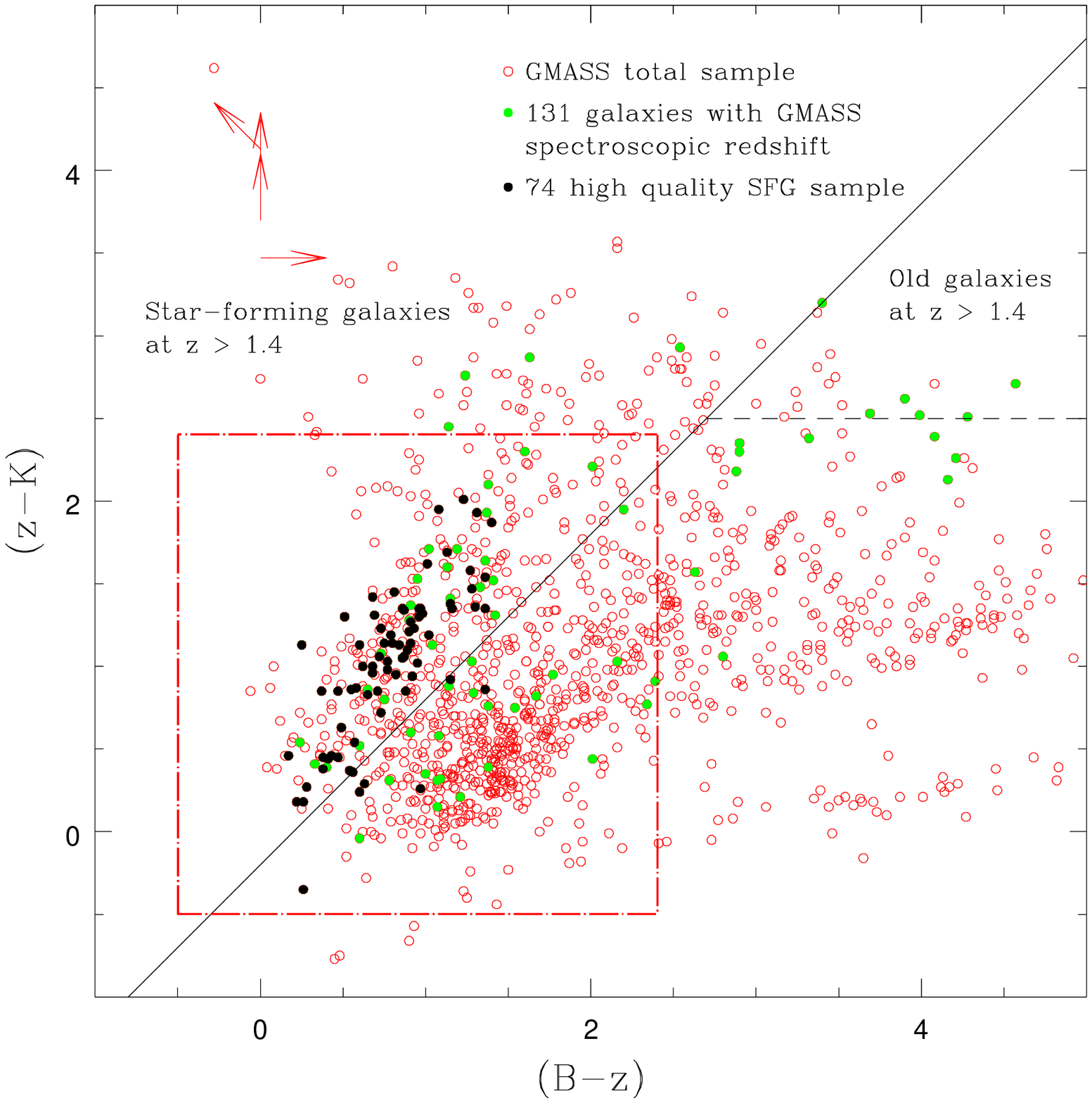}
\includegraphics[scale=0.4]{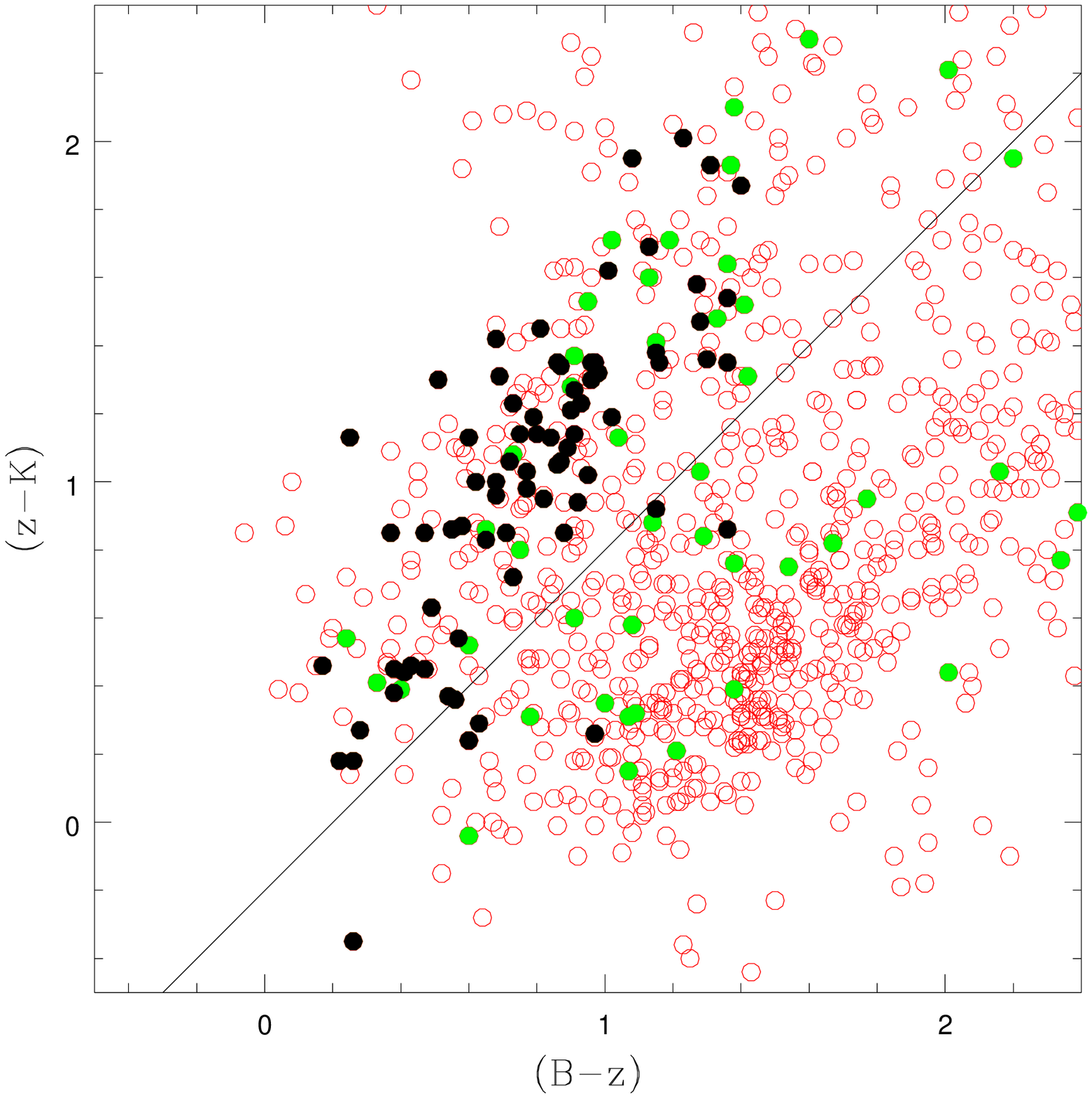}
\caption{Top panel: \emph{BzK} plot for the objects in the GMASS total sample (red open circles). The green dots represent the 131 objects with a secure redshift determination from the GMASS spectroscopic sample. The black dots represent the 74 high-quality SFGs analyzed in this work. Arrows indicate upper limits for objects undetected in one or more bands. Bottom panel: zoom on the 74 SFG sample.}
\label{BzK}
\end{figure}

\subsection{The GMASS survey}
\label{sec:The GMASS survey}

\begin{figure}[b]
\centering
\includegraphics[scale=0.4]{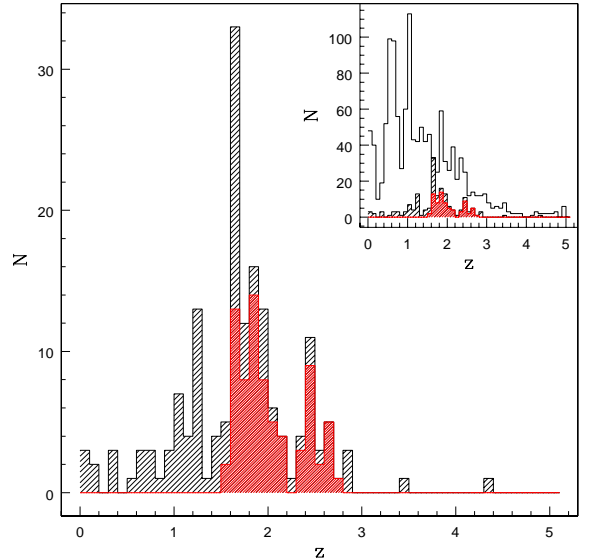}
\caption{Redshift distribution. The black shaded histogram represents secure spectroscopic redshifts of the observed full GMASS sample; the red shaded histogram represents redshifts for our sub-sample of 74 high-quality star-forming galaxies. In the inset, the distribution of photometric redshifts for the whole \emph{GMASS total sample} is also shown (empty histogram).}
\label{z}
\end{figure}
\begin{figure*}
\centering
\includegraphics[scale=0.4]{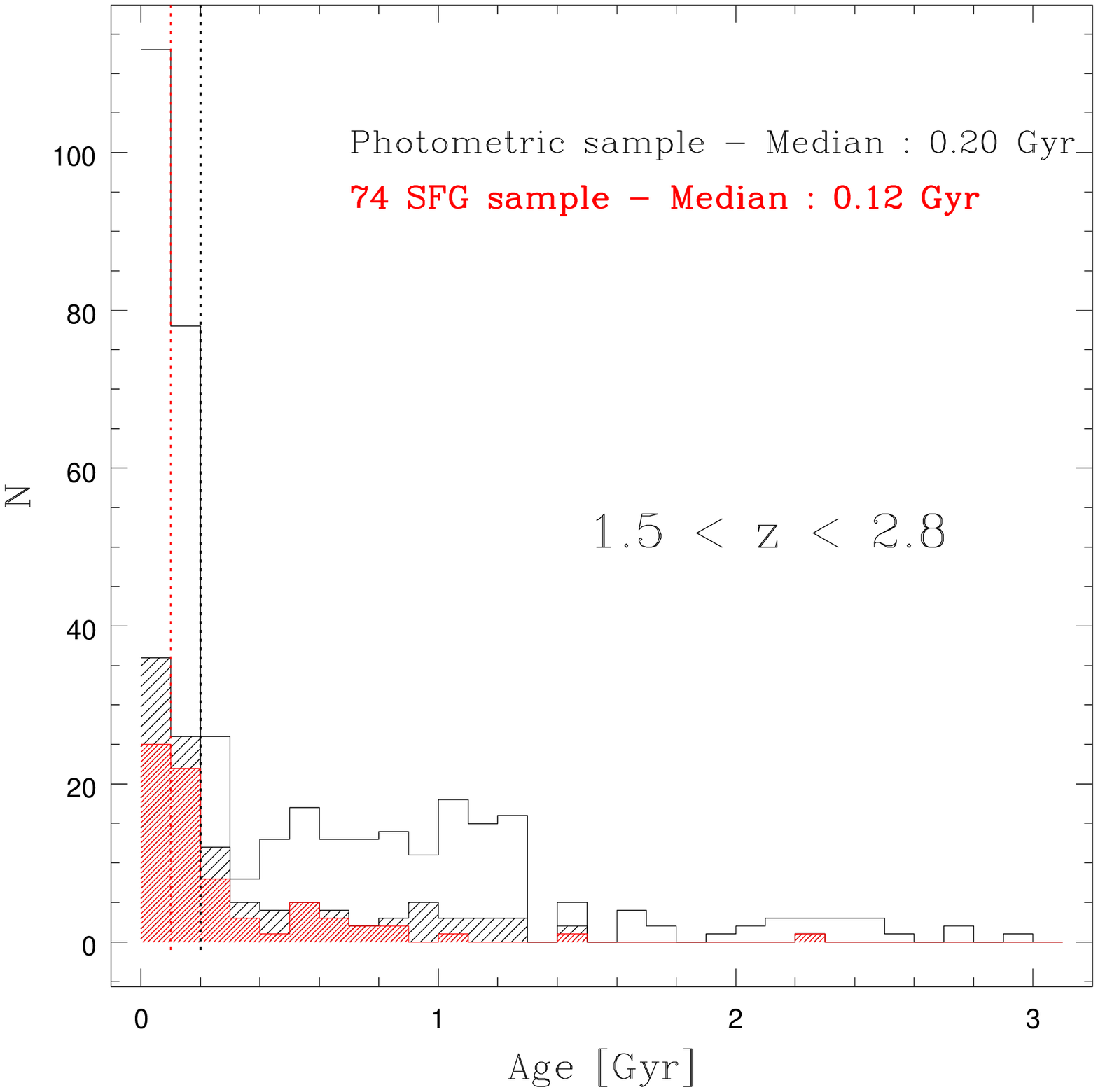}
\includegraphics[scale=0.4]{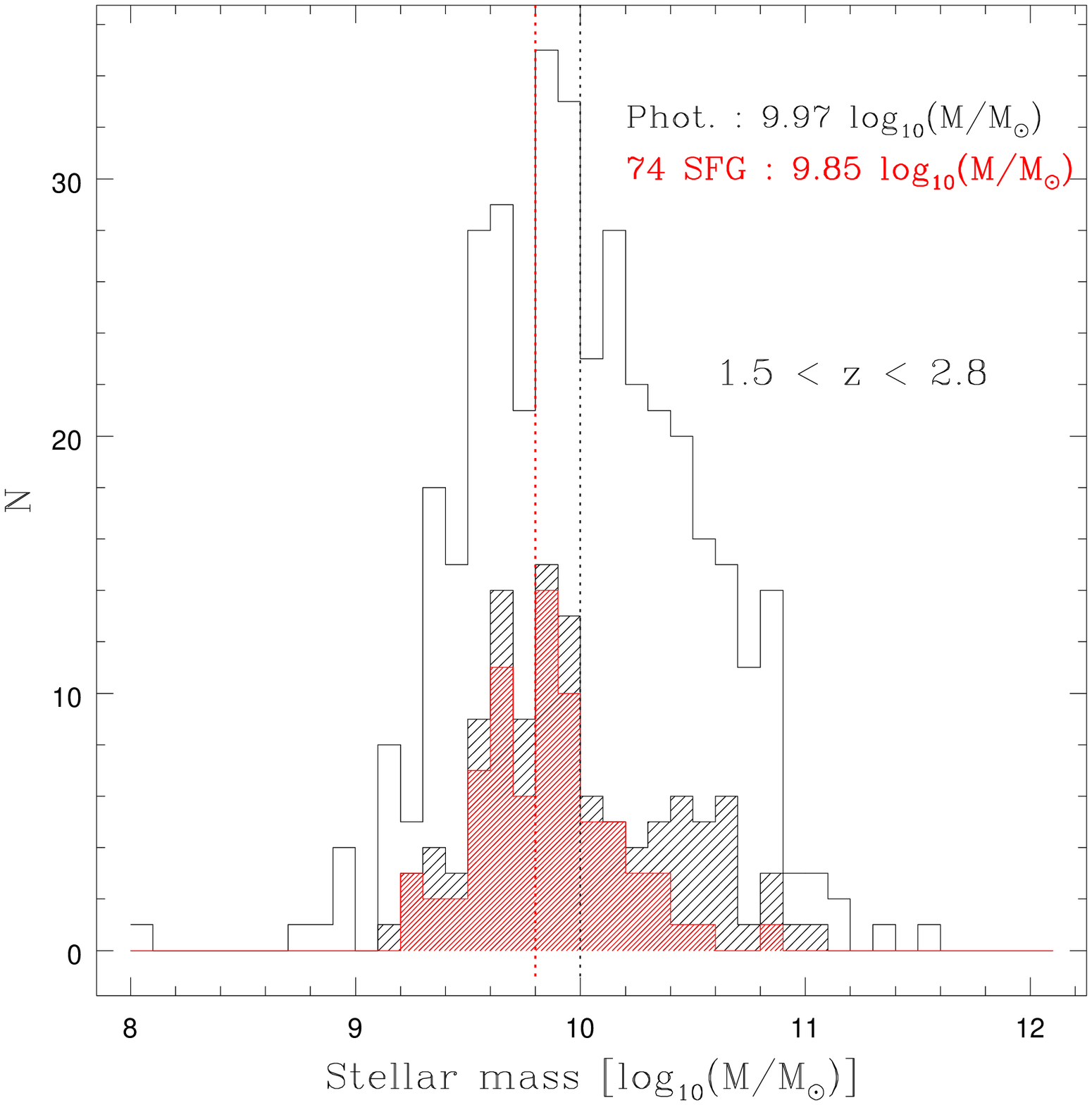}
\includegraphics[scale=0.4]{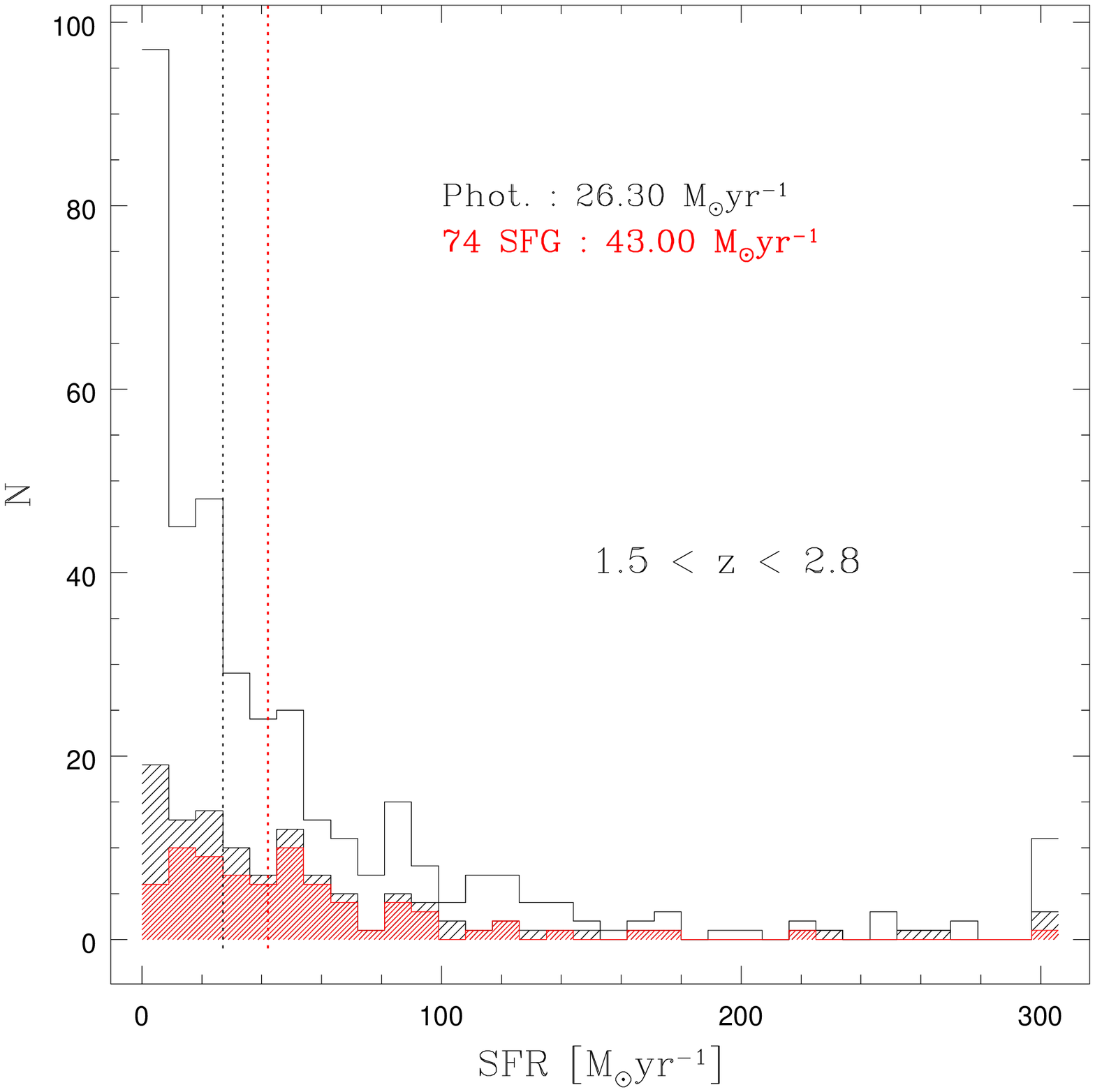}
\caption{Distributions, in the redshift range $1.5 < z < 2.8$, of stellar age, stellar mass and star-formation rate computed by SED fitting (using \citet{2005MNRAS.362..799M} models with a Kroupa IMF) for: the final sample of 74 objects analyzed in this paper (shaded red histogram); the GMASS total sample (open histogram); and objects with a secure redshift determination from the GMASS spectroscopic sample (shaded black histogram). Medians of the photometric catalogue and the 74 SFG sample are also shown (vertical black and red lines, respectively).}
\label{TOThist}
\end{figure*}

GMASS is an ESO VLT large program project based on data acquired using the FORS2 spectrograph. 
The project's main science driver is to use ultra-deep optical spectroscopy in order to measure the physical properties of galaxies at redshifts $1.5 < \emph{z} < 3$. 

The first step in the definition of the sample was to define a region of 6.8 $\times$ 6.8 arcmin$^2$ (matching the field of view of FORS2) in the GOODS-South public image taken at $4.5 \mu m$ with the \emph{Infrared Array Camera} (IRAC) mounted on the \emph{Spitzer Space Telescope}. 
The sample included all the sources with $\emph{m}_{4.5} < 23.0$ (AB system).

The choices of photometric band and limiting magnitude were imposed by several considerations. Among the IRAC bands, the $4.5 \mu m$ one offers the best compromise in terms of sensitivity, PSF, and image quality. In addition, it samples the rest-frame near-infrared up to $\emph{z} \sim 3$, thus allowing a selection that is more sensitive to stellar mass. This is also the band into which the rest-frame $1.6~\mu m$ peak of the stellar SEDs is redshifted for $\emph{z} > 1.5$. The limiting magnitude was mainly dictated by the observational constraints imposed by the spectrograph, in particular by the number of available slits with respect to the surface density of targets at $\emph{z} > 1.4$ detected in the field. Moreover, at $\emph{m}_{4.5} < 23.0$ the selection is sensitive to stellar masses down to log$(M/M_\odot) \approx 9.8$, 10.1, and 10.5 for $\emph{z} = 1.4$, 2, and 3, respectively. This ensures that it is possible to investigate the evolution of the galaxy mass assembly for a wide range of masses.
The photometric selection led to the construction of a catalogue of 1277 objects, the \emph{GMASS total sample}.
From this catalogue, 221 objects with photometric redshift $\emph{z}_{phot} \geq 1.4$ were selected as targets for ultra-deep spectroscopy. 
The total allocated observational time (145 hours) was distributed over six masks: three observed through the 300V grism (the blue masks) and three observed through the 300I grism (the red masks). 

Among the 221 objects, 170 could actually be put in the masks. To fill the masks, we also observed objects not included in the spectroscopic target list, obtaining a final \emph{GMASS spectroscopic sample} of 250 unique objects.
To ensure the feasibility of our spectroscopy, it was necessary to set magnitude limits B $<$ 26.0 and I $<$ 26.0 for the blue and red masks, respectively. The assignment of each object to either category was made according to colors and photometric redshifts: red objects at intermediate redshift ($(\emph{z} - \emph{K}) \geq 2.3$,~~$\emph{z}_{phot} \leq 2.5$) were listed as primary targets to be included in the red masks, while objects that are blue ($(\emph{z} - \emph{K}) < 2.3$) or have UV absorption lines that have been redshifted into the optical domain ($\emph{z}_{phot} > 2.5$) were allocated as primary targets for the blue masks. 
The actual wavelength coverage for a certain slit depends, apart from the grism and order separation filter, on its position in the mask in the dispersion direction. The maximum wavelength range covered for central slits was 3300-6500 $\AA$ for the blue masks and 6000-11000 $\AA$ for the red masks. To be able to include enough background for subtraction of the sky, minimum slit lengths were constrained to be 9$^{\prime\prime}$ and 8$^{\prime\prime}$ for the red and blue masks, respectively, while the slit width was always 1$^{\prime\prime}$ for both grisms. 
The spectral resolution of $\lambda/\Delta\lambda \approx 600$ was chosen because of the large wavelength coverage that it provides and is, however, high enough to resolve and identify spectral features for redshift determination. In total, it was possible to determine secure spectroscopic redshifts for 131 objects included in the spectroscopic target list, and for another 16 galaxies we obtained at least a redshift estimate. 

The physical properties of the galaxies belonging to the entire GMASS total sample were estimated by fitting different sets of evolutionary population synthesis models to broad-band SEDs (extending from \emph{U} band to the IRAC $8\mu m$ band) derived from photometric data. In particular, we used the results obtained with the synthetic spectra of \citet{2005MNRAS.362..799M}, adopting a Kroupa IMF \citep{2001MNRAS.322..231K} and fixed solar metallicity. Star-formation histories were parametrized by exponentials ($e^{\frac{-t}{\tau}}$) with e-folding timescales $\tau$ of between 100 Myr and 30 Gyr, plus the case of constant SFR. For the dust extinction, a Calzetti law \citep{2000ApJ...533..682C} was assumed. By means of a fitting procedure based on $\chi^{2}$ minimization, a best-fit model for each galaxy was derived, providing estimates for age, stellar mass, extinction $A_V$, and SFR \citep{2000A&A...363..476B, 2007A&A...474..443P}.

\subsection{Selection criteria and redshift refinement}
\label{sec:Selection criteria and redshift refinement}

\begin{figure}
\centering
\includegraphics[scale=0.4]{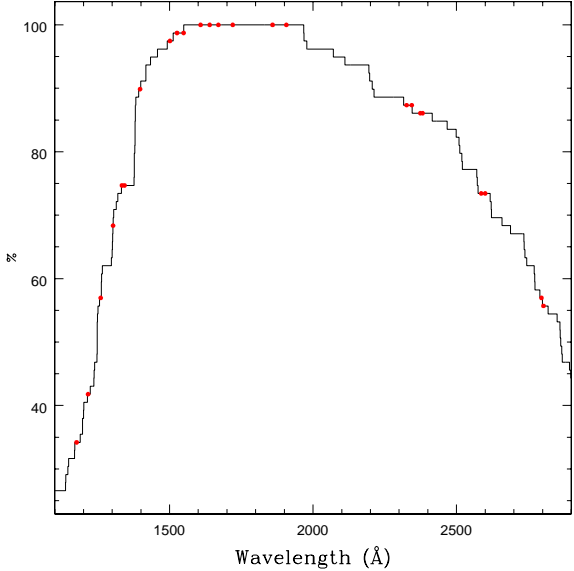}
\caption{Percentage of contributing objects per wavelength unit to the composite spectrum of the total sample (74 SFGs) (see Figure~\ref{aveTOTv2_label}). Red dots indicate the wavelength of the most prominent spectral lines detectable in the spectrum.}
\label{count}
\end{figure}

In line with the general aims of this work, we wished to select a representative sub-sample of SFG spectra from the GMASS sample, with confirmed spectroscopic redshift and high-quality spectra for rest-frame wavelengths between 1100 $\AA$ and 2900 $\AA$, which is both the range used by \citet{1994ApJ...429..582C} to define the continuum slope $\beta$, and the interval where we can find the blueshifted absorption spectral lines that are tracers of the warm phase of galactic-scale outflows. Therefore, we selected all the objects from the \emph{GMASS spectroscopic sample} that had been observed through a blue grism mask (B $<$ 26.0), and have a catalogue redshift $1.5 < \emph{z}_{spec} < 2.8$. These redshift and photometric criteria ensured the selection of star-forming galaxies. 
As an additional check, we plotted the selected objects on the colour-colour diagram developed by \citet{2004ApJ...617..746D}. 
The GMASS SFG sample analyzed in this work have been verified to lie in the star-forming region of the \emph{BzK} plot, as shown in Fig.\ref{BzK}. 
Since the spectral features related to star-formation activity may be "contaminated" by the presence of nuclear activity, objects with a spectrum showing AGN features were excluded. In particular, we found one object displaying broad emission lines, which was classified as a type 1 AGN, and another object with narrow C IV $\lambda1549~\AA$ emission typical of type 2 AGN. 
After these preliminary selection steps, we ended up with an ensemble of 94 SFGs.

The next selection was based on the quality of the redshifts. 
Since the redshifts have been originally obtained exploiting different measurement tecniques, we decided to homogenize them adopting a recursive procedure based on the cross-correlation of the single spectra with a template created from the sample itself \citep{1979AJ.....84.1511T}. 
Briefly, the peak of the cross-correlation function between the spectrum of a galaxy and a given template provides the redshift and velocity dispersion of the galaxy with respect to the template. To quantify the significance of the results, a cross-correlation goodness parameter R is defined.\footnote[2]{R is the ratio of the correlation peak height to the amplitude of the antisymmetric noise.}
The composite spectrum created by averaging the spectra from the preliminary sample of 94 SFGs, using catalogue redshifts, was taken as a template. From the cross-correlation of the template with each single galaxy spectrum, a new set of redshifts was obtained. On average, the difference between these new redshifts and the catalogue ones is $\Delta z \sim 2 \times 10^{-3}$. Then, the goodness parameter R was checked: too low a value of R means that only a low-quality redshift determination is possible because the spectrum is of low signal-to-noise ratio (S/N). After some tests,  we decided to use R $>$ 3 as the constraint determining that a redshift is reliable, and 20 objects, which did not satisfy this constraint, were excluded from the final high-quality sample.

The 74 remaining spectra were used to produce a new template, which was iteratively cross-correlated with each single spectrum until a round was reached when redshifts did not change significantly compared to the previous round. We reached convergence after five rounds and the final redshifts were the ones subsequently used in this work. Figure~\ref{z} shows the redshift distribution of the final sample of 74 high-quality SFGs, with reliable redshifts measured as explained above, against the original spectroscopic GMASS sample of 131 extragalactic objects. \citet{2009A&A...504..331K} found that the peak at $z \sim 1.6$ represents a galaxy overdensity, possibly a cluster under assembly, where a significant portion of galaxies have a passive spectrum and hence are not included in our SFG sample.

The distributions of age, stellar mass, and star-formation rate for the final sample of 74 SFGs, with respect to the GMASS total sample, are also shown in Figure~\ref{TOThist}, in the redshift range $1.5 < z < 2.8$. We note that the adopted selection criteria, especially the cut in optical magnitude, led our sample to be biased towards younger ages and higher star-formation rates, with respect to the parent galaxy population. 

\section{Rest-frame UV spectroscopic features}
\label{sec:Rest-frame UV spectroscopic features}

The composite spectrum of the sample of 74 SFGs is shown in Figure~\ref{aveTOTv2_label}\footnote[3]{The composite spectrum is publicly available at http://www.arcetri.astro.it/$\sim$cimatti/gmass/publicdata.html}. It was generated by stacking the individual spectra, shifted into their rest-frame using the redshifts defined in the previous section, and then normalized in the wavelength range 1600-1800$~\AA$ and rebinned to a dispersion of 1$~\AA$ per pixel\footnote[4]{The dispersion of single spectra is 2.5$~\AA$ in the observed frame. The rest-frame dispersion of the composite was set to be 1$~\AA \sim~ ~2.5~\AA~ / 1+\bar{z}~$, to match the observed pixel resolution.}. Figure~\ref{count} shows the percentage of contributing objects per wavelength unit to the composite spectrum, with an indication of the most prominent spectral lines used in our analysis. 
 
The continuum slope of the composite spectrum, measured following \citet{1994ApJ...429..582C}, is $\beta = -1.15 \pm 0.01$, which is in good agreement with the mean of the values measured for the 74 individual spectra in the sample: $<\beta>=-1.11 \pm 0.44$ (r.m.s.) (see the following data analysis section).  

Several types of emission and absorption lines are recognizable. The most prominent ones are the lines due to absorption, by the interstellar medium, of the radiation coming from stars, highlighted with red (low-ionization) and green (high-ionization) colours in Figure~\ref{aveTOTv2_label}. 

Information about the most prominent spectral lines detected in the spectrum is summarized in Table~\ref{tabLines}. 

\subsection{Photospheric absorption stellar lines}
\label{sec:Photospheric absorption stellar lines}

\begin{table*}
\caption[]{Spectral lines in a SFG rest-frame UV spectrum.}
\label{tabLines}
\centering                          
\begin{tabular}{l l l l}        
\hline\hline                 
Ion & Vacuum             & Rest-frame & Origin\\
    & wavelength ($\AA$) & EW\tablefootmark{a} ($\AA$) & \\
\hline                      
C III 			& 1175.71 & 1.70 	$\pm$ 0.12 		     & stellar (photospheric)\\
N V			& 1238.82, 1242.80  & -1.11	$\pm$ 	0.20    	    & ISM\\
Si II 			& 1260.42 & 1.47 	$\pm$ 0.09 		     & ISM\\
Si II*			& 1264.74 & -0.36       $\pm$ 0.25  		     & Fine-structure emission (ISM)\\    
O I - Si II\tablefootmark{b} 	& 1302.17, 1304.37 & 2.43 	$\pm$ 0.26\tablefootmark{d}  & ISM\\
Si II*			& 1309.28 & -1.26       $\pm$ 0.25  		     & Fine-structure emission (ISM)\\
C II 			& 1334.53 & 2.09 	$\pm$ 0.15  		     & ISM\\
O IV 			& 1343.35 & 0.27 	$\pm$ 0.07  		     & stellar (photospheric)\\
Si IV\tablefootmark{c}	& 1393.76 & 2.02 	$\pm$ 0.09  		     & ISM + stellar wind (P-Cygni profile)\\
Si IV\tablefootmark{c}	& 1402.77 & 1.42 	$\pm$ 0.05  		     & ISM + stellar wind (P-Cygni profile)\\
Si III			& 1417.24 & 0.72 	$\pm$ 0.29    		     & stellar (photospheric)\\
S V 			& 1501.76 & 0.54 	$\pm$ 0.13  		     & stellar (photospheric)\\
Si II 			& 1526.71 & 1.94 	$\pm$ 0.08  		     & ISM\\
C IV\tablefootmark{b,c}	& 1548.20, 1550.78 & 3.24 	$\pm$ 0.16\tablefootmark{d}  & ISM + stellar wind (P-Cygni profile)\\
Fe II 			& 1608.45 & 1.31 	$\pm$ 0.11  		     & ISM\\
He II			& 1640.42 & -1.02 	$\pm$ 0.10  		     & stellar wind\\
Al II 			& 1670.79 & 1.54 	$\pm$ 0.08  		     & ISM\\
Ni II			& 1709.60 & 0.70	$\pm$ 0.35    		     & ISM\\  
N IV			& 1718.55 & 1.14 	$\pm$ 0.18    		     & stellar wind (P-Cygni profile)\\
Ni II			& 1741.55 & 0.25	$\pm$ 0.41    		     & ISM\\  
Ni II			& 1751.91 & 0.39	$\pm$ 0.41    		     & ISM\\  
Si II			& 1808.01 & 0.76	$\pm$ 0.09    		     & ISM\\  
Si I			& 1845.52 & 0.53	$\pm$ 0.21    		     & ISM\\  
Al III\tablefootmark{c}	& 1854.72 & 1.20 	$\pm$ 0.21  		     & ISM + stellar wind (P-Cygni profile)\\
Al III\tablefootmark{c}	& 1862.79 & 0.85 	$\pm$ 0.11 		     & ISM + stellar wind (P-Cygni profile)\\
C III]\tablefootmark{b,c}	& 1906.68, 1908.68 & -1.02 	$\pm$ 0.15\tablefootmark{d}  & nebular\\
Fe II			& 2249.88 & 0.60	$\pm$ 0.22    		     & ISM\\  
Fe II			& 2260.78 & 0.50	$\pm$ 0.31    		     & ISM\\
C III			& 2297.58 & 0.28 	$\pm$ 0.24    		     & stellar (photospheric)\\     
C II]   		& 2326.00 & -1.24 	$\pm$ 0.20    		     & nebular\\
Fe II 			& 2344.21 & 2.45	$\pm$ 0.13    		     & ISM\\
Fe II*			& 2365.66 & -0.74  	$\pm$ 0.38  		     & Fine-structure emission (ISM)\\    
Fe II\tablefootmark{c} 	& 2374.46 & 1.87	$\pm$ 0.13    		     & ISM\\
Fe II\tablefootmark{c} 	& 2382.76 & 2.22	$\pm$ 0.15    		     & ISM\\
Fe II*			& 2396.15 & -1.12  	$\pm$ 0.18  		     & Fine-structure emission (ISM)\\    
Fe II\tablefootmark{c} 	& 2586.65 & 2.65	$\pm$ 0.17    		     & ISM\tablefootmark{e}\\
Fe II\tablefootmark{c} 	& 2600.17 & 2.77	$\pm$ 0.14    		     & ISM\tablefootmark{e}\\
Fe II*			& 2612.65 & -1.07  	$\pm$ 1.35    		     & Fine-structure emission (ISM)\\    
Fe II*			& 2626.45 & -1.17  	$\pm$ 0.41    		     & Fine-structure emission (ISM)\\    
Mg II\tablefootmark{c} 	& 2796.35 & 2.43	$\pm$ 0.10    		     & ISM\tablefootmark{f}\\
Mg II\tablefootmark{c} 	& 2803.53 & 2.07	$\pm$ 0.10    		     & ISM\tablefootmark{f}\\
Mg I			& 2852.96 & 1.75	$\pm$ 0.68    		     & ISM\\	
\hline\hline
\end{tabular}
\tablefoot{
\tablefoottext{a}EWs measured in the composite spectrum of 74 SFGs. Tha values are negative for emission lines and positive for absorption lines.\\
\tablefoottext{b}Blended at our spectral resolution.\\
\tablefoottext{c}Doublet.\\
\tablefoottext{d}The measure of the EW refers to the blend of the two lines.\\
\tablefoottext{e}Possible stellar wind contribution.\\
\tablefoottext{f}Possible photospheric stellar contribution.\\
}
\end{table*}

Photospheric stellar lines generate by the absorption of radiation produced in the stellar interiors by the most external gas layers. Being very weak, these lines are almost undetectable in the single spectra of our sample, because of the low S/N in the continuum (typical $<$5), but they can be both identified and measured in composite spectra, since the averaging procedure improves the S/N ($>$15) (see Figure~\ref{sing_ave}). 
As discussed later in more detail, stellar photospheric lines can be used to define the velocity zero point of the system, with respect to which the velocities of the outflowing interstellar gas are calculated.  

In our study here, we were able to identify five stellar lines: C III $\lambda1176~\AA$, O IV $\lambda1343~\AA$, S III $\lambda1417~\AA$, S V $\lambda1500~\AA$, and C III $\lambda2297~\AA$. 

\subsection{Nebular emission lines}
\label{sec:Nebular emission lines}

\begin{figure*}
\centering
\includegraphics[trim=13 0 0 300, clip=true, width=6.5in]{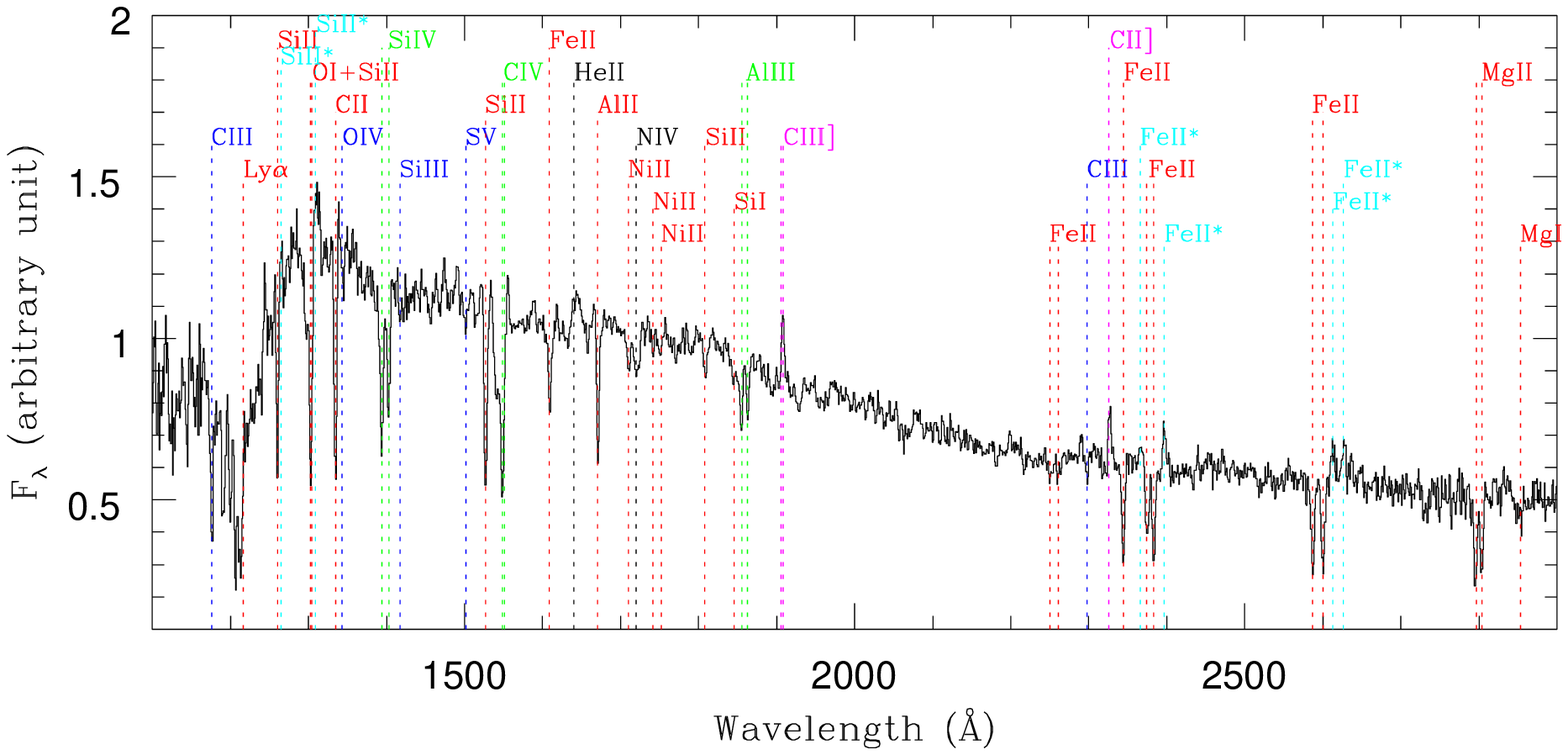}
\includegraphics[scale=0.78]{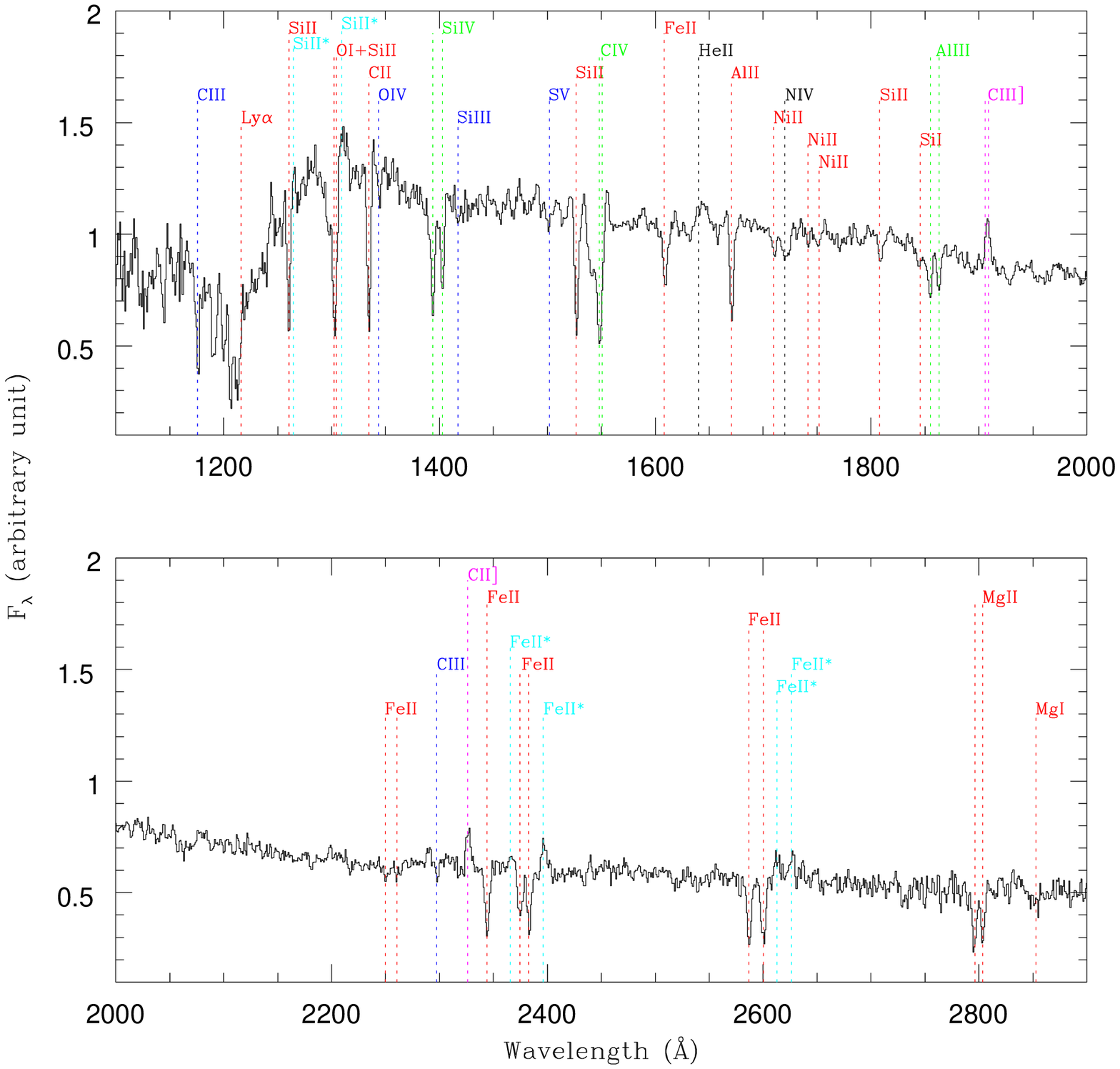}
\caption{Top panel: average spectrum of the total sample (74 SFGs). Bottom panels: zoom of the composite spectrum, above and below 2000 $~\AA$. Spectral lines of interest are labeled. Blue: absorption stellar photospheric lines; red: interstellar absorption low-ionization lines; green: interstellar absorption high-ionization lines; magenta: emission nebular lines; black: emission and absorption lines associated with stellar winds; cyan: interstellar fine-structure emission lines.}
\label{aveTOTv2_label}
\end{figure*}

Two semi-forbidden carbon lines are detectable in our spectra: C III] $\lambda1909~\AA$ and C II] $\lambda2326~\AA$, unresolved at our spectral resolution. These lines originate in nebular regions photoionized by radiation from massive O and B stars, hence can in principle be used, as well as photospheric stellar lines, to determine the system rest-frame velocity.

The C III] $\lambda1909~\AA$ is known to be a doublet ($\lambda\lambda1906.08,1908.73~\AA$). The flux intensity ratio of the two lines depends on the physical conditions of the emitting gas, in particular the electron density \citep{1992ApJ...389..443K}. For the electronic densities of local star-forming regions the ratio varies between 0.5 and 1.5, with 1.5 being the most typical value and therefore the one we imposed in the line measurements.

\subsection{Low- and high-ionization absorption lines associated with the ISM}
\label{sec:Low- and high-ionization absorption lines associated with the ISM}

\begin{figure*}
\centering
\includegraphics[scale=0.3]{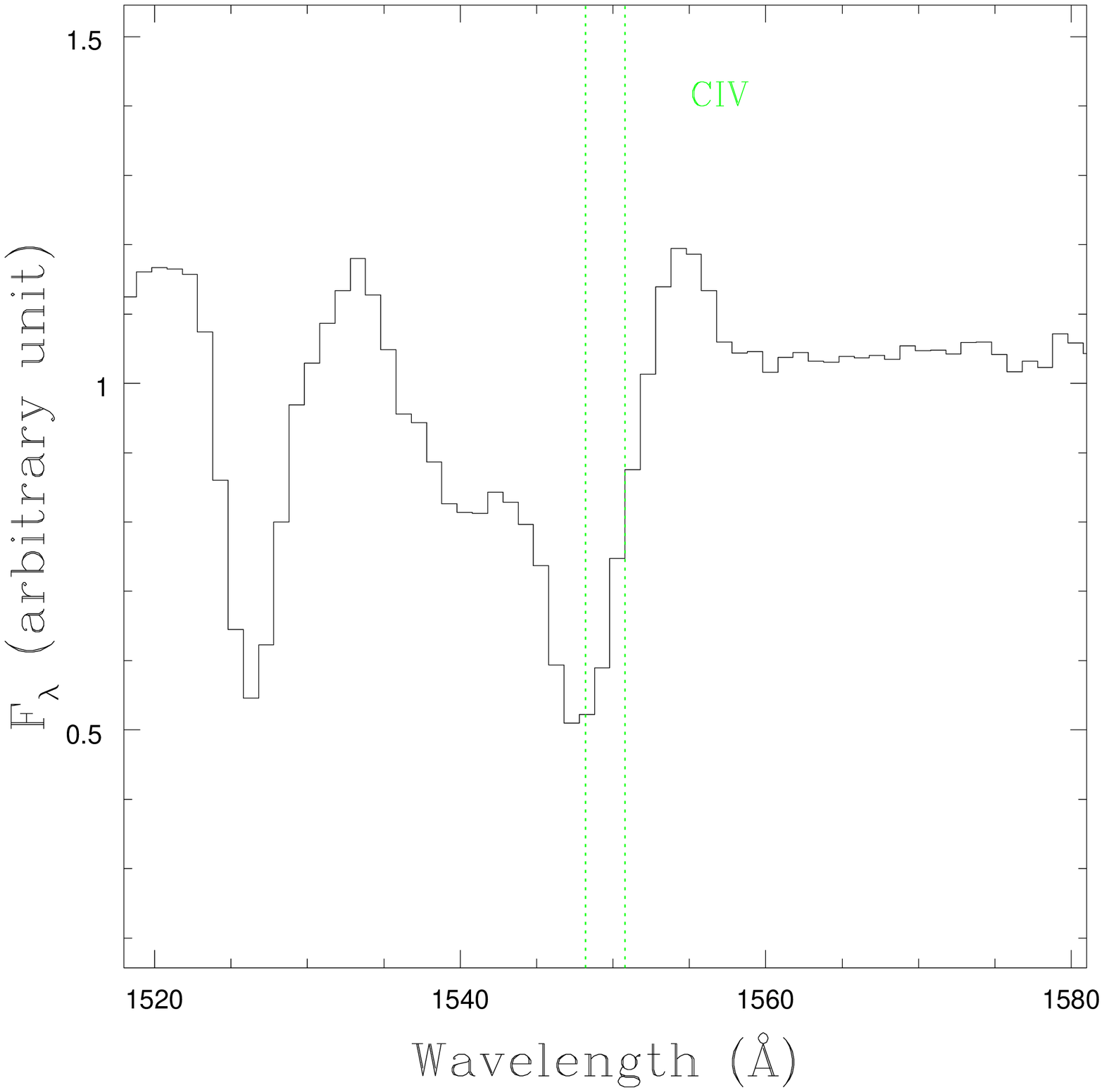}
\includegraphics[scale=0.3]{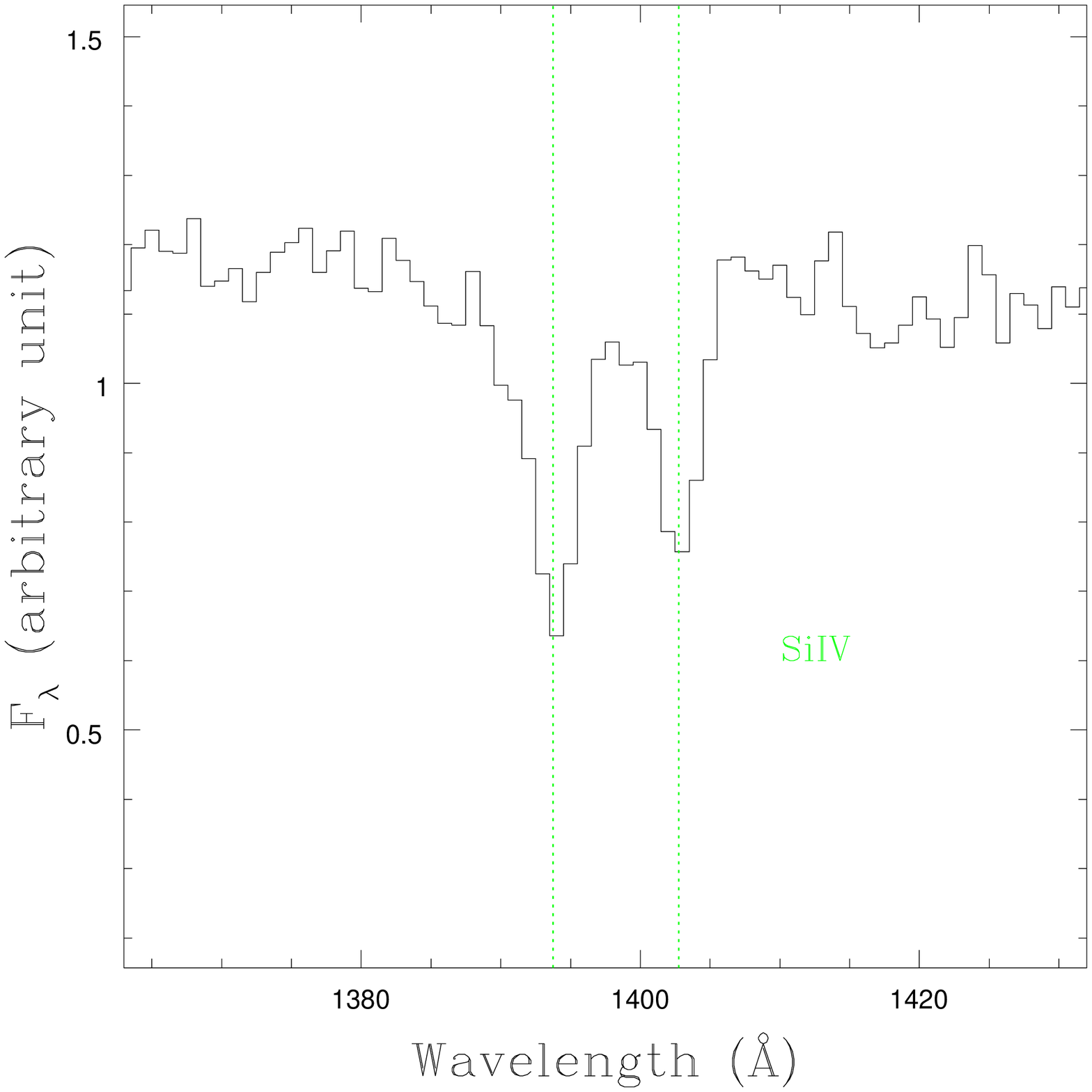}
\caption{Composite spectrum of the 74 spectra of the SFG sample (see Figure~\ref{aveTOTv2_label}): zoom-in of the regions aroud the C IV and the Si IV doublets. The spectrum has been shifted into its stellar rest-frame. The vertical lines indicate the vacuum wavelength of the line.}
\label{CIV_SiIV}
\end{figure*}

The strongest lines detectable in the UV rest-frame spectrum of starburst galaxies are low- and high-ionization absorption interstellar lines. 
The first set of lines are associated with neutral gas regions and caused by the absorption of stellar radiation by the interstellar medium, which ensures that they are the optimal tools for detecting outflows in the host galaxy. 

In particular, the strongest lines in our spectra are Si II $\lambda1260~\AA$, C II $\lambda1334~\AA$, Si II $\lambda1526~\AA$, Fe II $\lambda1608~\AA$ and Al III $\lambda1670~\AA$, Fe II $\lambda2344~\AA$, Fe II $\lambda2374~\AA$, Fe II $\lambda2382~\AA$, Fe II $\lambda2586~\AA$, Fe II $\lambda2600~\AA$, plus the blend (at our spectral resolution) of O and Si II at $\lambda1303~\AA$, and the doublet Mg II $\lambda\lambda2796,2803~\AA$. Some other weaker lines, which we identified as interstellar absorption lines, can be found in Table~\ref{tabLines}.

High-ionization lines predominantly trace gas with T $>10^4 K$. Three doublets are recognizible in our spectra, i.e., Si IV $\lambda\lambda1393,1402~\AA$, C IV $\lambda\lambda1548,1550~\AA$, and Al III $\lambda\lambda1854,1862~\AA$. Figure~\ref{CIV_SiIV} shows zoomed-in spectral regions of the composite spectrum, containing the CIV and SiIV doublets. The spectrum has been shifted into its stellar rest-frame and the expected position of the spectral lines is indicated by the dotted vertical lines.
Unlike low-ionization lines, more than one process causes the formation of the high-ionization doublets, including absorption of stellar radiation by the interstellar medium, as well as the contributions of stellar winds and photospheric stellar absorption, but these different components are difficult to distinguish.
A study of UV O-type stellar spectra by \citet{1984ApJ...280L..27W} qualitatively showed that the Si IV doublet profile is dominated by the interstellar component, because only blue giant and supergiant stars contribute to the wind component, while the C IV doublet is dominated by the stellar wind feature, whose contribution to the line profile comes from stars belonging to all luminosity classes, from main-sequence to supergiant stars. 
The stellar wind component should display a P-Cygni profile, which is characterized by redshifted emission and blueshifted absorption. Since spectral resolution is too low to allow a decomposition of the line profile in its componentes, high-ionization lines are poorly suited to the search and study of large-scale outflows.

\subsection{Other notable spectral lines}
\label{sec:Other notable spectral lines}

\begin{figure*}
\centering
\includegraphics[trim=13 0 0 300, clip=true, width=6.5in]{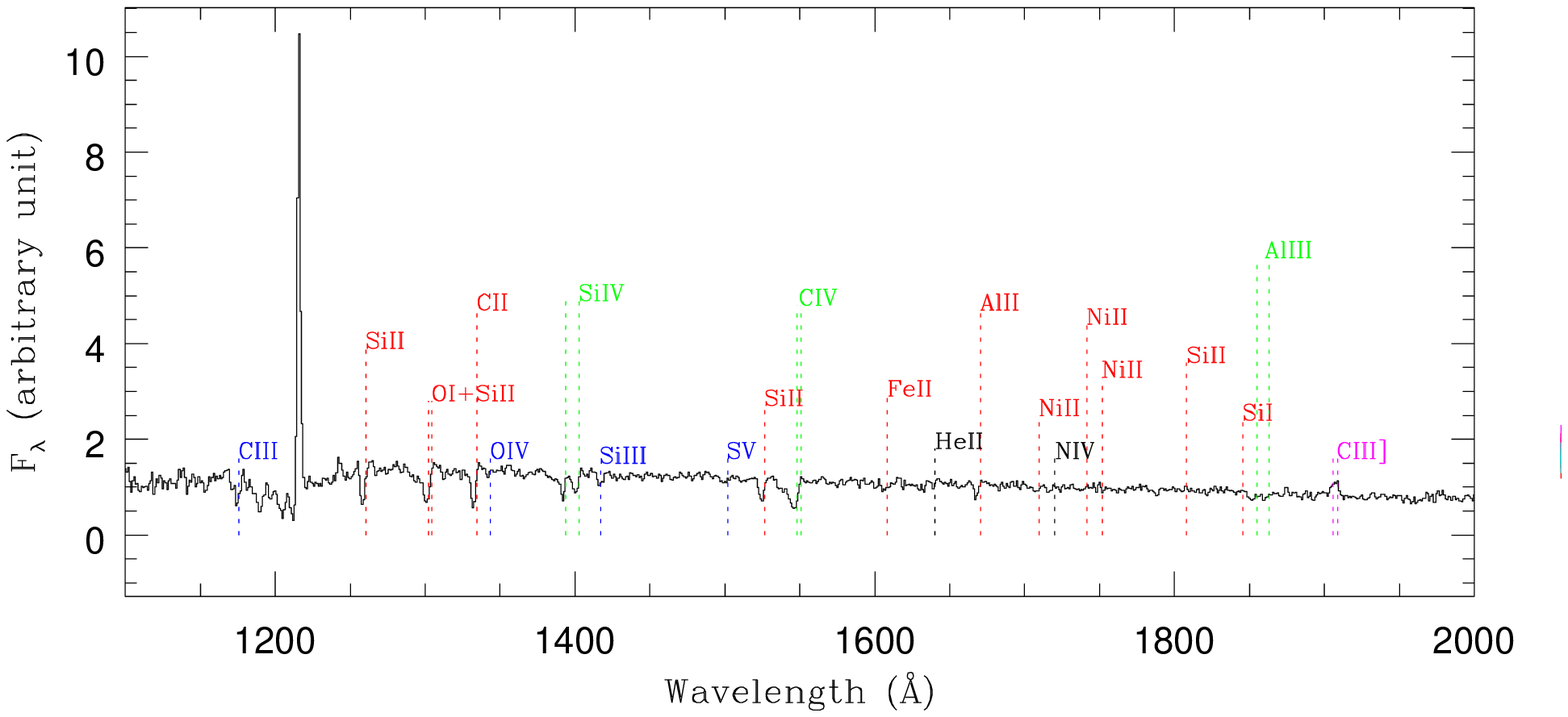}
\caption{Average spectrum of the six galaxies with Ly$\alpha$ line in emission. The spectrum has been shifted into the rest-frame given by the Ly$\alpha$. The expected position of the most prominent spectral lines is indicated. See Figure~\ref{aveTOTv2_label} for the colour legend.}
\label{aveLya6}
\end{figure*}

\subsubsection{Ly$\alpha$}

The Ly$\alpha$ $\lambda1216~\AA$ spectral line originates from recombinations in H II regions and can be used to study the physical conditions within them. When detected in emission, it is the most prominent spectral feature in the spectra of high-redshift galaxies. When seen in absorption, the line has a very broad profile, often blueshifted. 
In our SFG sample, 32 spectra extend blueward to include the Ly$\alpha$ region: 8 of them show the line in emission, while 24 display absorption. We measured the velocity offset between the Ly$\alpha$ in emission and the low-ionization interstellar absortion lines in the six spectra where both sets are detectable. A mean velocity offset $\Delta\emph{v}_{em-abs} \sim 566 \pm 27$ km/s was found, consistent with the $\sim$ 650 km/s reported by \citet{2003ApJ...588...65S} for their sample of $\emph{z} \sim 3$ Lyman break galaxies.

Two composite spectra were then created, one with the spectra with Ly$\alpha$ in emission (see Figure~\ref{aveLya6}) and the other with the spectra with Ly$\alpha$ in absorption, and their continuum slope was measured following the procedure described in a following section. We found the emission Ly$\alpha$ spectrum to have a bluer slope with respect to the other case (respectively, $\beta_{emLy\alpha} = -1.40 \pm 0.06$ and $\beta_{absLy\alpha} = -1.04 \pm 0.03$), thus confirming the relation between UV continuum slope, and therefore dust extinction, and the Ly$\alpha$ profile in SFGs, as already observed in LBGs at $z \sim 3$.

\subsubsection{He II $\lambda1640~\AA$}

He II $\lambda1640~\AA$ is a typical feature produced by winds from Wolf-Rayet (WR) stars or AGN.
WR stars are thought to represent a stage in the evolution of O-type stars with masses  $\emph{M} > 20 M_\odot$. They are characterized by strong winds with velocities reaching thousands km/s, causing considerably mass losses. The strength of the broad He II $\lambda1640~\AA$ is likely to depend on the ratio of WR with respect to O-type stars, which is in turn strongly determined by the metallicity \citep{1998ApJ...497..618S, 1995A&A...298..767M, brinchmann2008}: at lower metallicities O-type stars are less likely to evolve into WR, and the He II becomes intrinsecally weaker. From the EW of He II $\lambda1640~\AA$ measured on the composite spectrum of our entire sample, $EW_{HeII} = 1.02 \pm 0.10~\AA$, we derived a WR/(WR+O) ratio of $0.05 \pm 0.008$ \citep{1998ApJ...497..618S}. 

Quite a good agreement is found with the models of \citet{brinchmann2008}. For the case of nearly constant SFR, with an age of $\sim 120$ Myrs, which is the median age of our sample of galaxies, the predicted EW values range from $\sim 0.65$ to $\sim 1.3~\AA$ \citep[see][Figure~2]{brinchmann2008}, depending on the metallicity ($0.4 < Z/Z_\odot < 1.0$). Since the metallicity of the SFGs in the GMASS sample has been estimated to be $Z \sim 0.3 Z_\odot$ by \citet{halliday2008}, our measured EW is fairly consistent with the model predictions.

Most recent spectral population synthesis codes include also the modeling of massive binaries system \citep{eldridge2009, brinchmann2008bis}, which may significantly effect the evolution of the WR stars. Up to now these models have only be tested in the local Universe, but their success in reproducing the observed spectral features also looks promising with regard to their application to higher redshift samples.

\subsubsection{Fe II*}

In our composite spectrum, we were also able to identify some iron fine-structure emission lines, between 2000$\AA$ and 3000$\AA$: Fe II* $\lambda2365~\AA$, Fe II* $\lambda2396~\AA$, Fe II* $\lambda2612~\AA$, and Fe II* $\lambda2626~\AA$. These lines have been proposed to be generated by photon scattering in large-scale outflows \citep{2010arXiv1008.3397R}, therefore their detection can be considered as a confirmation of the presence of outflows.

\section{The UV continuum slope and dust extinction}
\label{sec:The UV continuum slope and dust extinction}

\begin{figure}
\centering
\includegraphics[scale=0.4]{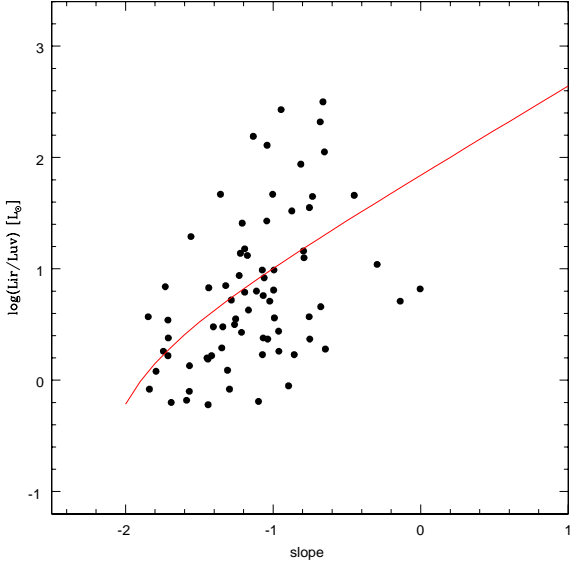}
\caption{UV continuum slope as a function of the ratio between UV luminosity (not extinction corrected) and IR luminosity. The red line shows the \citet{1999ApJ...521...64M} relation.}
\label{meurer}
\end{figure}

Under the assumption, which is quite robust when considering galaxies dominated by a young stellar population, that the shape of the UV continuum can be fairly accurately approximated by a power law
\\
\\$F_\lambda \varpropto \lambda^\beta$,
\\
\\where $F_\lambda$ is the observed flux ($erg~s^{-1}~cm^{-2}~\AA^{-1}$) and $\beta$ is the continuum slope, \citet{1994ApJ...429..582C} showed that there is a strong correlation between the slope $\beta$ and nebular extinction measured in the optical using the Balmer decrement.

\begin{figure}
\centering
\includegraphics[scale=0.4]{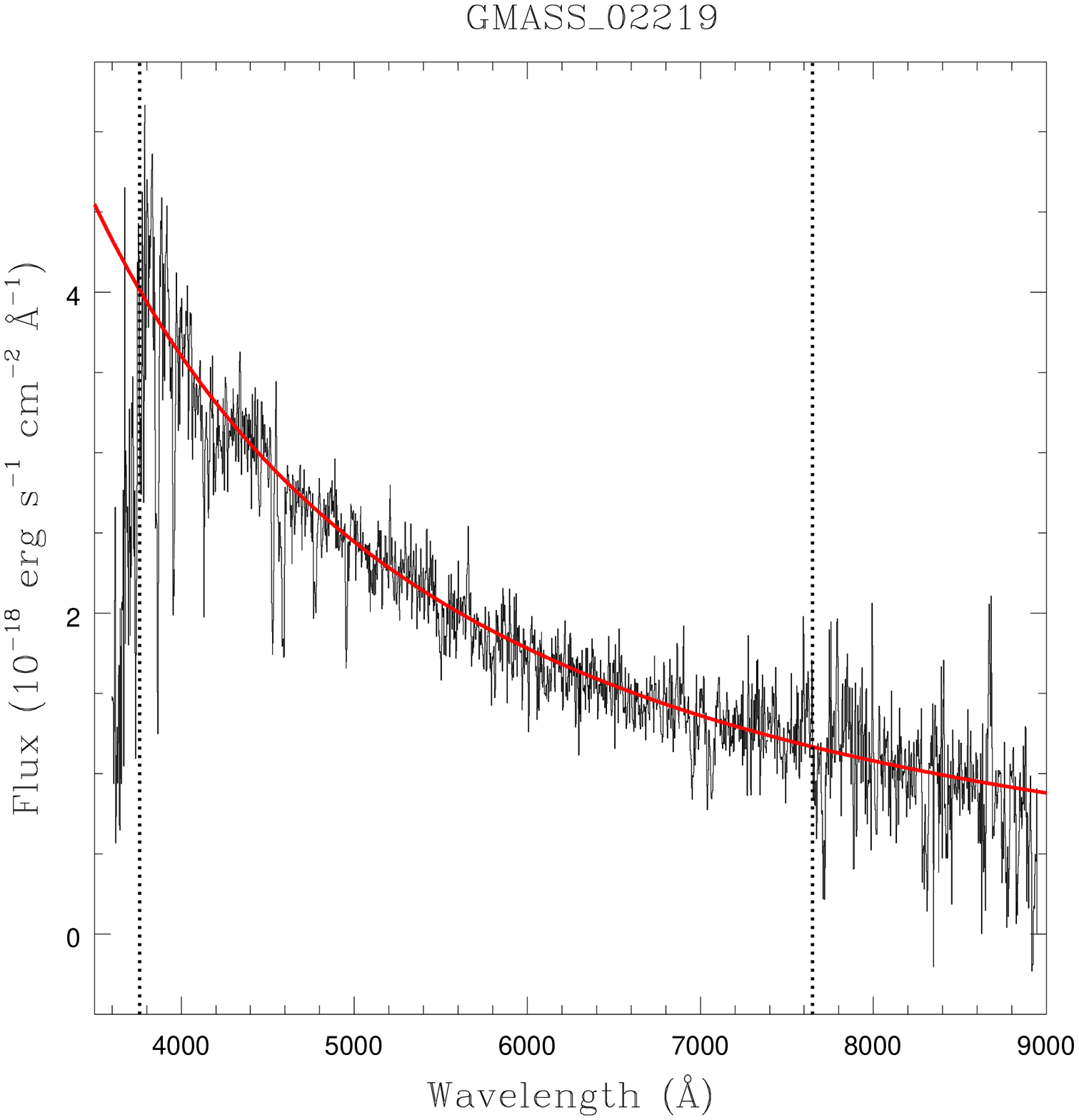}
\includegraphics[scale=0.4]{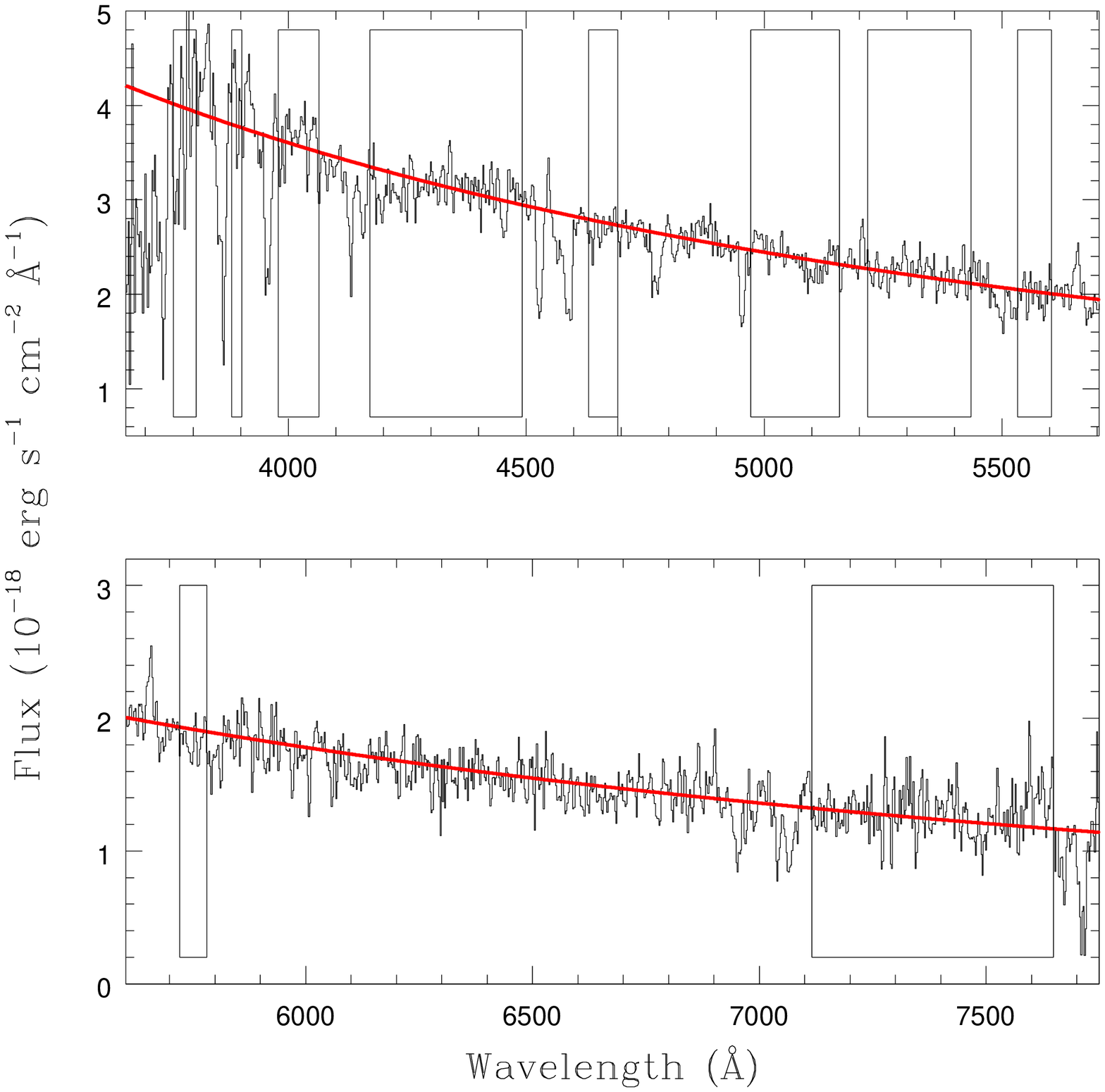}
\caption{Top panel: a spectrum from our sample with corresponding fit of the continuum superimposed in red. The vertical dashed lines delineate the range covered by the windows defined by \citet{1994ApJ...429..582C} and used in the fitting procedure. Bottom panels: zoom over the wavelength range used in the fitting routine, with the ten fitting-windows indicated.}
\label{fit}
\end{figure}

\begin{figure}
\centering
\includegraphics[scale=0.4]{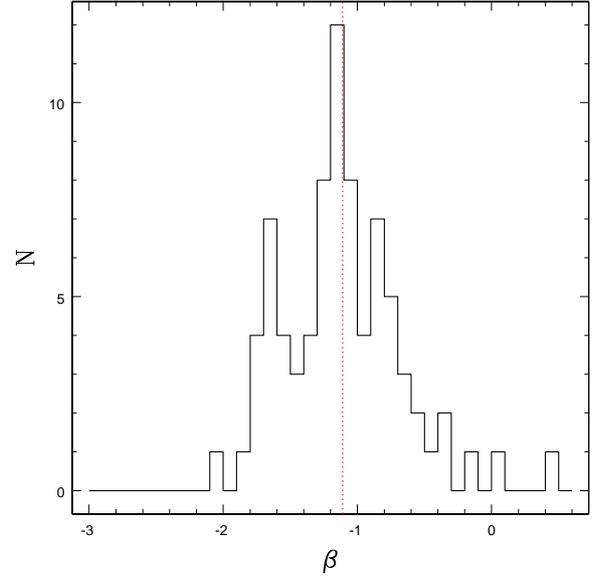}
\caption{UV continuum slope distribution over the sample of 74 SFGs. The vertical red lines indicate the mean value for the sample: $<\beta>=-1.11$, with $r.m.s. = 0.44$.}
\label{hist_slope}
\end{figure}

For each of the 74 spectra in our sample of high-quality SFGs, the continuum was fitted with a least squares minimization, in the $log(F_\lambda)~-~log(\lambda)$ plane, of the average fluxes measured in ten wavelength windows in the range between $\sim 1250~\AA$ and $2600~\AA$, as defined by \citet{1994ApJ...429..582C}. These windows are chosen to avoid the strongest spectral features.
In Figure~\ref{fit}, we present a spectrum from our sample, as an example. The range covered by the fitting windows is indicated by the vertical dotted lines in the upper panel, while the final fit is superimposed on the spectrum in red. The two lower panels show zoomed-in regions of the same spectrum, where the ten fitting windows (in the $F_\lambda~-~\lambda$ plane) are indicated.
The UV continuum slope ($\beta$) distribution over the sample can be seen in Figure~\ref{hist_slope}. The mean value for the sample is $<\beta>=-1.11 \pm 0.44$ (r.m.s.).

The definition of $\beta$ is not unique across the literature. First of all, it can be calculated directly on spectra, as we did, or determined from the photometry \citep{2009ApJ...705..936B}. Then, the range over which $\beta$ is defined also changes from one work to another: when defined over a shorter wavelength range than that defined by \citet{1994ApJ...429..582C}, the quoted slope is systematically redder \citep[see][]{2001PASP..113.1449C}.

Keeping all this in mind, we compared the results of our analysis with others, for both local and high redshift galaxies, taken from the literature. 
The mean value of $\beta$ over our sample is in good agreement with the $<\beta>~=-1.13 \pm 0.12$ measured by \citet{1998ApJ...503..646H} from the spectra of 45 local star-forming galaxies observed with the \emph{IUE} satellite. 
For high redshift galaxies, \citet{2005A&A...444..137N} analyzed the spectra of 34 star-forming galaxies at $2 < z < 2.5$ from the FORS Deep Field (FDF) spectroscopic survey \citep{2004A&A...418..885N}. The determined mean value of the continuum slope for the sub-sample of galaxies with bluer spectra is $<\beta>~=-1.01 \pm 0.11$, with which the mean value over our sample is fairly consistent, taking into account the different wavelength baseline ($\sim 1200 - 1800~\AA$ in their analysis, compared to $\sim 1200 - 2600~\AA$ in this work).
Finally, \citet{2009ApJ...705..936B} presented a systematic measurement of the UV continuum slope over a wide range in redshift, using photometry. For galaxies at redshift $z \sim 2$, they find $<\beta>~ \sim -1.5$, with a scatter of $\sim 0.30$ \citep[see][Figure~5]{2009ApJ...705..936B}, with which our results agree reasonably well, again considering the differences in methods and wavelength baselines used. 
\\ 
\\

\citet{1999ApJ...521...64M} found an empirical relation between the ultraviolet spectral slope and dust extinction at 1600$~\AA$, in local star-forming galaxies given by
\\
\\$A_{1600} = 4.43 + 1.99~\beta$.
\\
\\
We used this relation to estimate dust extinction for the galaxies of our sample. 

The \citet{1999ApJ...521...64M} relation was derived from an observed correlation between the ratio of infra-red luminosity over ultra-violet luminosity (not corrected for extinction) and the continuum slope. We compared the continuum slopes computed for our sample with the ratio between luminosities measured at $1500 \AA$ (see next section) and the IR luminosities extrapolated from the Spitzer-MIPS 24 $\mu m$ fluxes (following \citet{2007ApJ...670..156D}), and our galaxies were found to be in reasonable agreement with the \citet{1999ApJ...521...64M} relation, as it is shown in Figure~\ref{meurer}.

Using the $A_{1600}$ values derived from the continuum slope, and a Calzetti law \citep{2000ApJ...533..682C}, the color excess was derived. This determination of $E(B-V)$ was then compared with other estimates: the extinction obtained by SED fitting and the values calculated from B and z bands using the relation of \citet{2004ApJ...617..746D}, proposed to be valid for star-forming galaxies in the redshift range $1.4 < z < 2.5$
\\
\\$E(B-V)~=~0.25~(B~-~z~+~0.1)$,
\\
\\where \emph{B} and \emph{z} are AB magnitudes.

\begin{figure}
\centering
\includegraphics[scale=0.4]{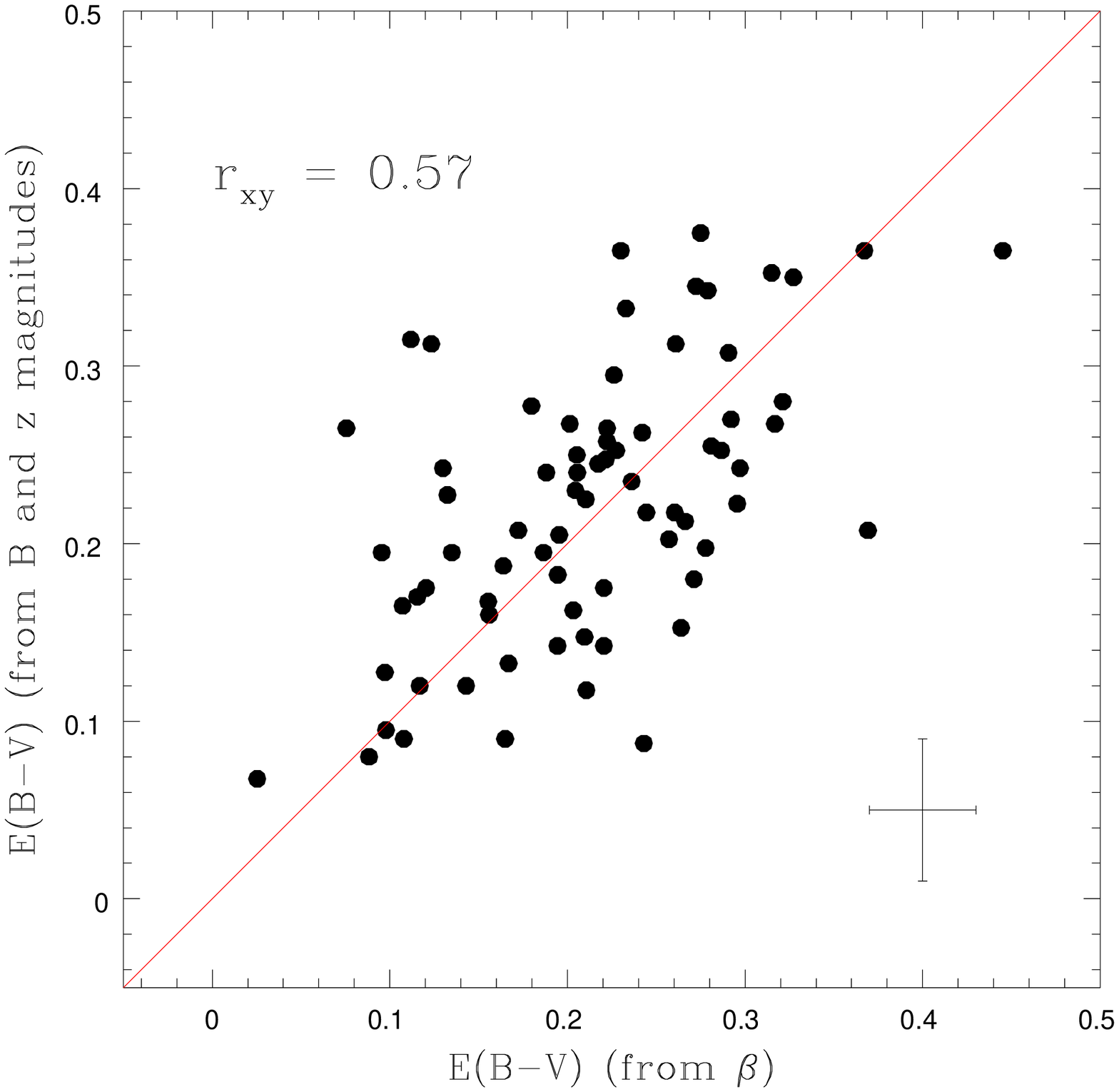}
\includegraphics[scale=0.4]{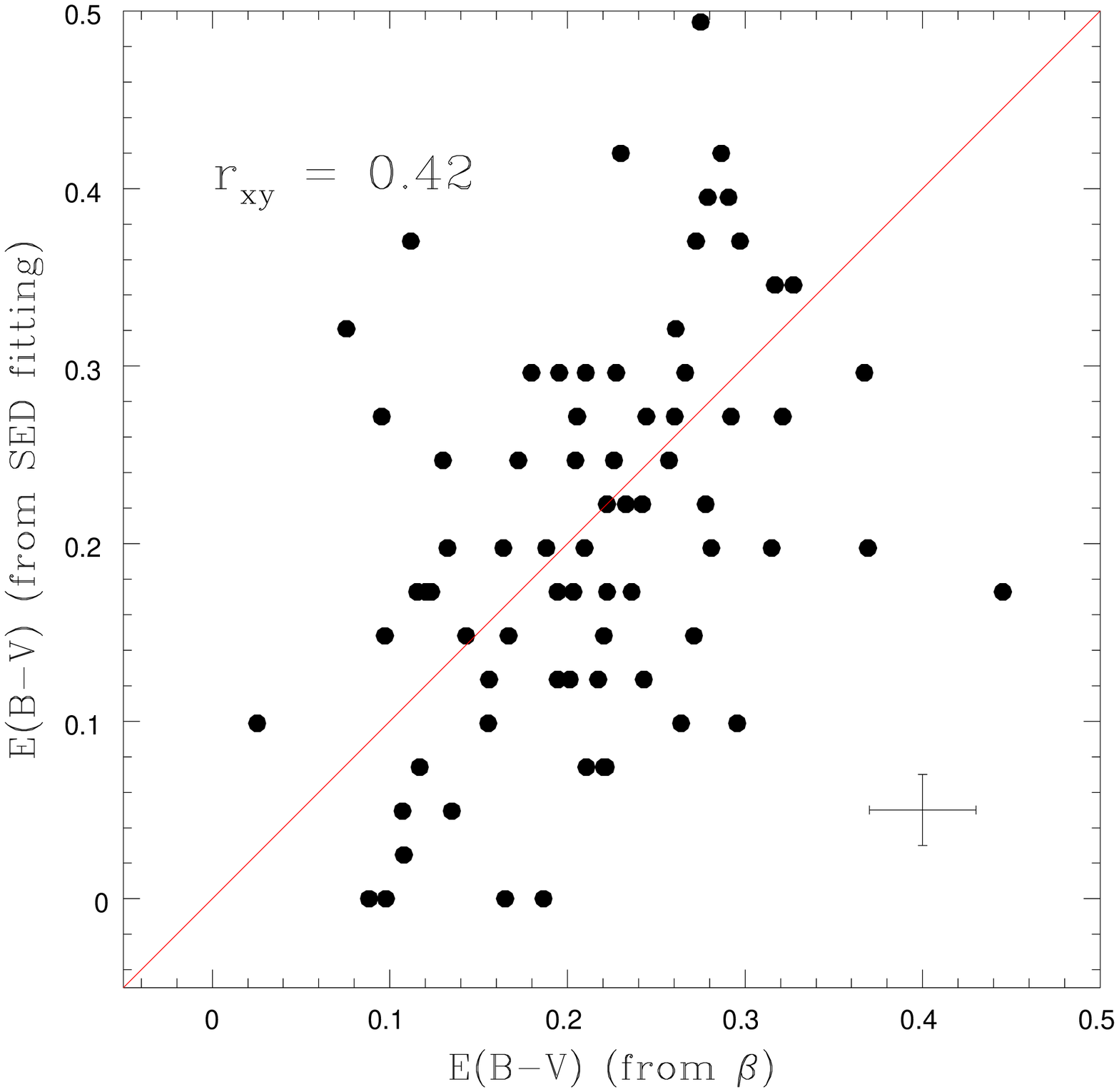}
\caption{Comparison between E(B-V) derived from spectra UV continuum slope and, respectively, E(B-V) calculated from B and z magnitudes (top panel) and E(B-V) from SED fitting (bottom panel), with the one-to-one line in red. The Pearson correlation coefficient $r_{xy}$ is also reported.}
\label{ebv}
\end{figure}

Figure~\ref{ebv} compares the different estimates. The color excess from the continuum slope is quite consistent with the estimate derived from B and z bands, while it shows a larger dispertion when compared to the estimate from SED fitting. However, in both cases the probability that the two compared data sets are linearly uncorrelated, as derived by the Pearson coefficients ($r_{xy} = 0.57$ and $r_{xy} = 0.42$), is $<$ 0.1\%.
 
\section{Star-formation rate}
\label{sec:Star-formation rate}

\begin{figure}
\centering
\includegraphics[scale=0.4]{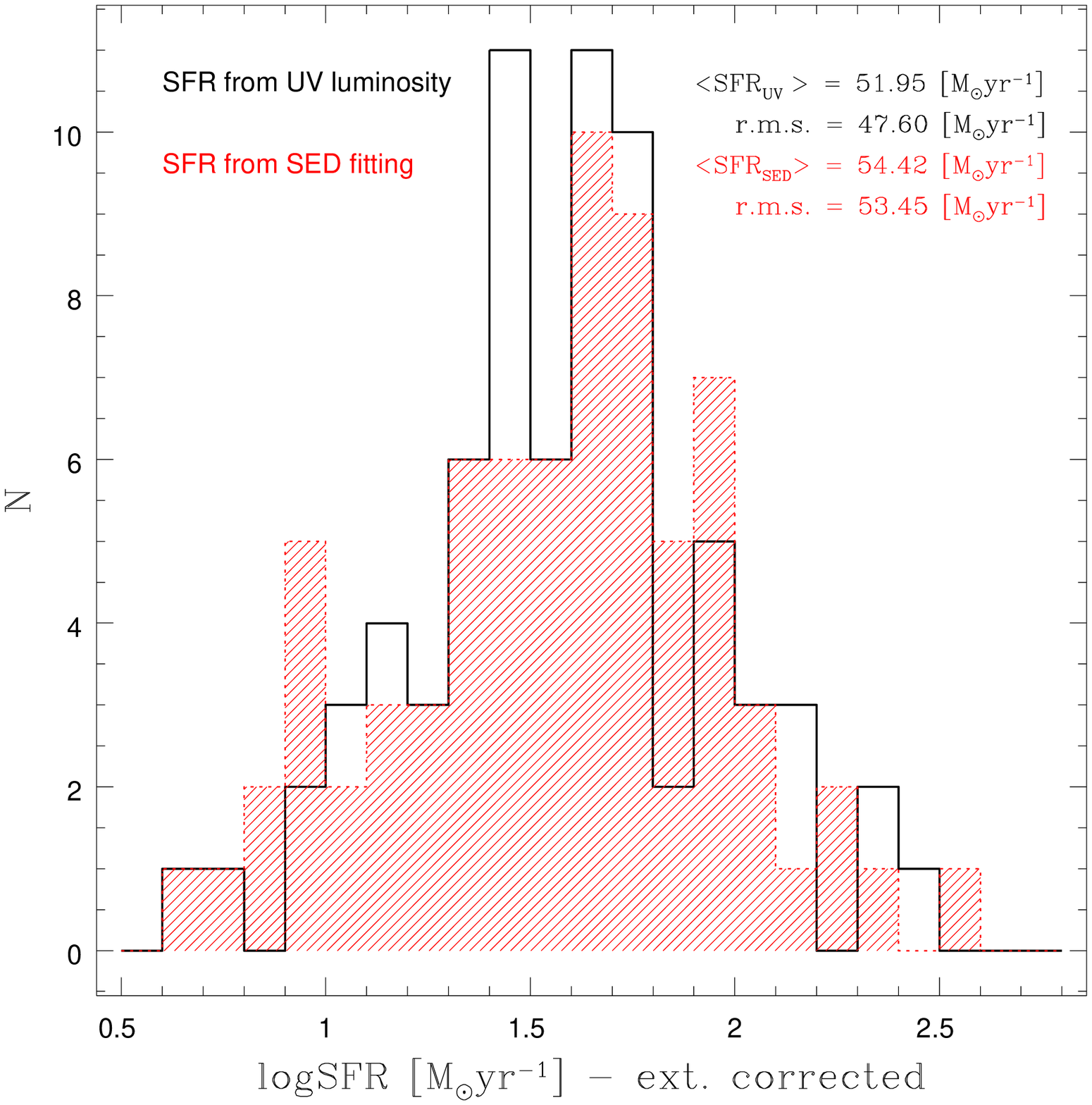}
\caption{Distributions of extinction-corrected SFR [$M_\odot~ yr^{-1}$] derived from UV luminosity measured on spectra and extinction-corrected SFR [$M_\odot~ yr^{-1}$] from SED fitting (using \citet{2005MNRAS.362..799M} models with a Kroupa IMF).}
\label{SFR}
\end{figure}

After estimating the dust extinction from the continuum slope of the spectra in our galaxy sample, the star-formation rates from the rest-frame UV luminosity were derived, by applying the relation from \citet{1998ARA&A..36..189K} scaled for a Kroupa IMF
\\
\\$SFR_{UV}~ (M_\odot~ yr^{-1}) = 0.9 \times 10^{-28} \times L_{1500~\AA}~ (erg~ s^{-1}~ Hz^{-1})$.
\\
\\First, the spectra were corrected for slit losses, so that integrated fluxes calculated for specific bandpasses would match catalogue magnitudes obtained by photometry. The photometric values were found to be on average, about one magnitude brighter than the ones inferred from spectra.
We checked whether a correlation existed between $\Delta$mag (defined as $mag_{phot} - mag_{spec}$) and the photometric magnitude, but found none. No correlation was found between $\Delta$mag and the morphology of the objects. The absence of correlations between the magnitude corrections and galaxy properties likely indicates that the main cause of light loss in the spectral data is the slit width not encompassing the full seeing disk.

For each of the corrected spectra, the 1500$~\AA$ luminosity was obtained by measuring the average flux in a wavelength window ranging from 1350~$~\AA$ to 1650$~\AA$. Unextincted fluxes were then calculated using, for each galaxy, the estimate of dust extinction previously derived from its continuum slope. 

It is worth noting that an $R_V$ value had to be chosen to convert the estimated extinction into color excess, and we used the one from \citep{2000ApJ...533..682C}. But to correct the UV fluxes we directly used the $A_{1600}$ estimated from the UV continuum slope. Therefore, our extinction correction does not depend on the choice of a specific extinction law.

The distribution of SFRs estimated from spectra, along with the ones computed by SED fitting, is presented in Figure~\ref{SFR}. 
Throughout this paper, as pointed out in section \ref{sec:The GMASS survey}, all the SED fitting derived quantities were computed using \citet{2005MNRAS.362..799M} models with a Kroupa IMF. These models were chosen because they include a treatment of the thermally pulsing asymptotic giant branch (TP-AGB) phase of stellar evolution, but the SED fitting procedure was also performed using \citet{2003MNRAS.344.1000B} models with a \citet{2003PASP..115..763C} IMF \citep[see][]{2008A&A...482...21C}. We compared the SFR obtained from UV luminosity with both the estimates from SED fitting, and we found the values computed with \citet{2005MNRAS.362..799M} models to be the more consistent with the spectra-based estimate.

The galaxies of our sample show average  $<SFR>~=52 ~M_\odot~ yr^{-1}$, with $r.m.s. = 48 ~M_\odot~ yr^{-1}$, which is similar to the findings of \citet{2007ApJ...670..156D} for their sample of K-selected star-forming galaxies at $z \sim 2$, and of \citet{2003ApJ...588...65S} for their sample of LBGs at $z \sim 3$.
\subsection{The SFR-stellar mass correlation}
\label{sec:The SFR-stellar mass correlation}

\begin{figure*}
\centering
\includegraphics[trim=0 0 0 80, clip=true, scale=0.7]{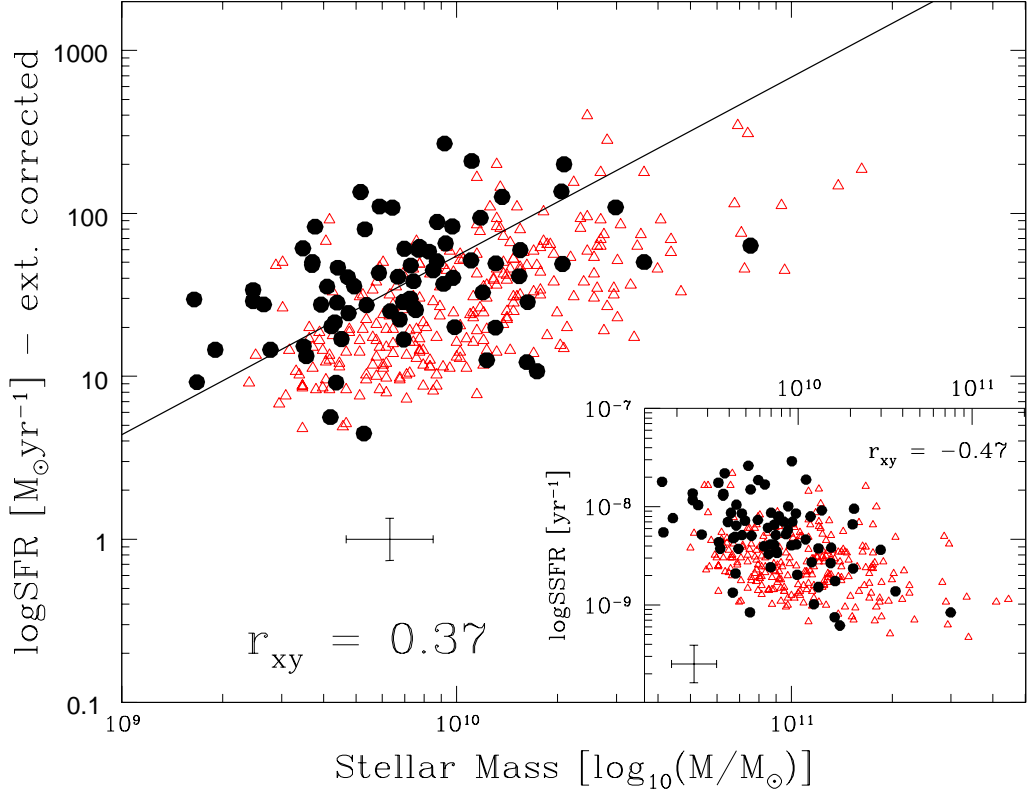}
\caption{Stellar mass as a function of SFR (big panel) for the GMASS sample analyzed in this paper (black dots). Stellar masses have been computed by means of SED fitting (using \citet{2005MNRAS.362..799M} models with a Kroupa IMF). The extinction-corrected SFR [$M_\odot~ yr^{-1}$] was derived from UV luminosity measured from spectra. The black line is a linear fit to the GMASS data (in logarithmic scale). We also indicate the Pearson correlation coefficient $r_{xy}$. The open red triangles represent the sample of $z \sim 2$ star-forming galaxies from \citet{2007ApJ...670..156D} (SFRs and masses scaled for a Kroupa IMF). In the little panel, stellar mass as a function of SSFR is shown.}
\label{SFRvsMASS}
\end{figure*}

The existence of a correlation between stellar mass and SFR was also investigated, and the results are shown in Figure~\ref{SFRvsMASS}. 
A positive correlation between SFR and stellar mass has been reported to hold at different redshifts \citep{2007A&A...468...33E, 2007ApJ...660L..43N, 2007ApJ...670..156D, 2009A&A...504..751S, 2009ApJ...698L.116P, 2010A&A...518L..25R}, despite the different techniques of SFR estimates used (radio luminosity and UV luminosity alone or combined with $\emph{L}_{IR}$), but the debate about its general validity and implications remains open. 
For our sample, we also found that more massive galaxies have higher SFRs (as estimated from $L_{1500~\AA}$), in agreement with previous studies. The big panel of Figure~\ref{SFRvsMASS} shows the stellar mass-SFR correlation for our spectroscopic sample of galaxies and for the photometric sample from \citet{2007ApJ...670..156D} (for the sake of consistency, we scaled the values from \citet{2007ApJ...670..156D} for our chosen IMF, before comparing the results). 

In order to estimate the slope of the SFR-Mass relation, a regression technique which treats symmetrically the variables is required. In a case where the intrinsic scatter is likely to dominate any errors arising from the measurement process, the unweighted ordinary least-squares (OLS) bisector has been demonstrated to be a recommended choice \citep{1990ApJ...364..104I,1992ApJ...397...55F,2011ApJ...728...72F}. In this method there is however the implicit assumption that the errors in the two quantities must be of the same order of magnitude. In our study, we derived statistical errors on the SFR estimated from the spectra, and derived from simulations the errors on masses estimated using SED fitting techniques, and we found that both errors are on average about 30$\%$, therefore the OLS bisector regression method is applicable.
In Figure~\ref{SFRvsMASS} the black line, that comes from the linear fit to the data (on a logarithmic scale) of the GMASS sample, has a slope of $1.10 \pm 0.10$, which is in agreement with the relations found for other $z \sim 2$ samples, in particular \citet{2007ApJ...670..156D}, who quote a slope of $\sim 0.9$ using UV-derived SFRs, and \citet{2009ApJ...698L.116P}, who report a slope $0.95 \pm 0.07$ and whose SFRs were computed from radio luminosity. We note that a similar proportionality is also seen at lower redshift \citep{2007A&A...468...33E, 2007ApJ...660L..43N}.

We also found that, in our sample, the SSFR decreases with the stellar mass (see Figure~\ref{SFRvsMASS}, insert panel), a trend with which the data from \citet{2007ApJ...670..156D} also seem to be consistent. A SSFR that declines with increasing stellar mass is found also in \citet{2010A&A...518L..25R}, for $z \sim 2$ sources, while \citet{2009ApJ...698L.116P} and \citet{2009MNRAS.394....3D} report in their analyses that the SSFR of $z \sim 2$ star-forming galaxies is almost independent of stellar mass. 

However, we recall that several selection effects and biases might affect our sample of 74 SFGs. A more extended analysis of the SFR at $z \sim 2$ will be the presented in a forthcoming paper, for a larger sample of galaxies and using IR data provided by Herschel. 

\section{Outflows}
\label{sec:Outflows}

\begin{figure}
\centering
\includegraphics[scale=0.4]{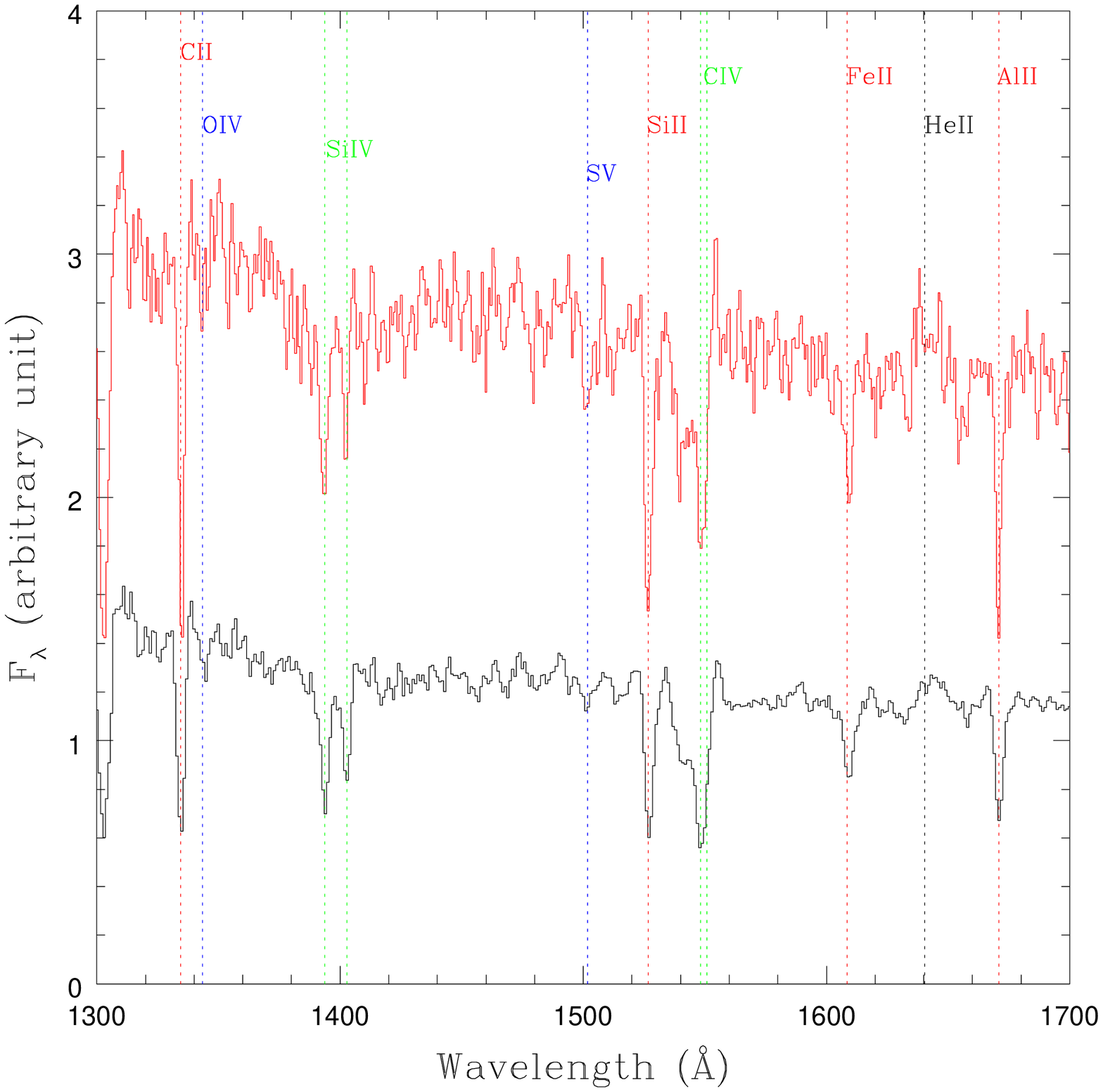}
\caption{Comparison of the total average spectrum (black, below; the same seen in Figure~\ref{aveTOTv2_label}) of our sample with the spectrum of a single galaxy (red, above). See Figure~\ref{aveTOTv2_label} for the colour code adopted.}
\label{sing_ave}
\end{figure}

We search for galactic-scale outflows in galaxy UV spectra by measuring the wavelength shift, of the centroid position, of spectral lines that originate from the absorption of the background stellar radiation by the interstellar medium. The most accurate estimate of the systemic rest-frame velocity of a galaxy is the one determined from spectral lines that are of stellar origin. The assumption that stellar spectral features are at rest with respect to their host galaxy is justified by considering that it is not possible to resolve stellar populations in high-redshift galaxies, and that the effects of stellar proper motions on a spectral line shift tend to statistically cancel out. The stellar lines observed in a galaxy UV rest-frame are extremely weak, and in the single GMASS spectra are almost undetectable because of the low S/N of the stellar continuum. It is possible to improve the S/N to identify the stellar spectral lines by working on composite spectra (see Figure~\ref{sing_ave}).

\begin{figure*}
\centering
\includegraphics[trim=0 0 0 180, clip=true, width=6.5in]{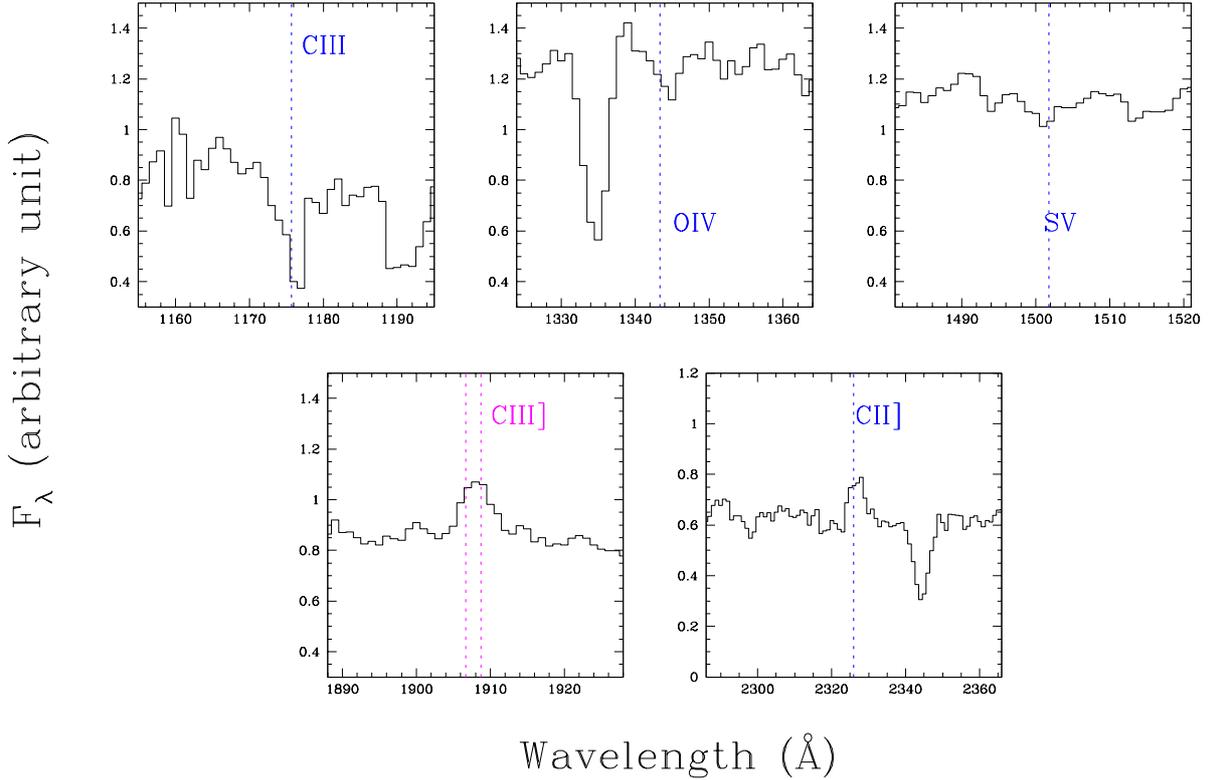}
\caption{Composite spectrum of the 74 spectra of the SFG sample (see Figure~\ref{aveTOTv2_label}): zoomed image of the three detectable stellar photospheric absorption lines (C III $\lambda1176~\AA$, O IV $\lambda1343~\AA$, S V $\lambda1501~\AA$. The latter was not used in the analysis since it is too weak to be reliable measured), the C III]  $\lambda\lambda1906,1908~\AA$ nebular emission line doublet and the CII] $\lambda2326~\AA$ nebular line. The vertical lines indicate the vacuum wavelength of the line.}
\label{prof_stell}
\end{figure*}

In Figure~\ref{aveTOTv2_label}, the composite spectrum of 74 high-quality star-forming galaxies is shown, where two stellar and two nebular lines can be detected and reliably measured: the absorption lines C III $\lambda1176~\AA$, O IV $\lambda1343~\AA$, and the semi-forbidden C III]  $\lambda\lambda1906,1908~\AA$ doublet and CII] $\lambda2326~\AA$ emission line. Since the redshifts of the single galaxies are determined from the strongest lines, which are the interstellar absorption ones, the composite spectrum does not appear to be at rest with respect to the stellar and nebular features (see Figure~\ref{prof_stell}). Therefore, in order to compute the velocity offsets of the lines tracers of the IS gas, first of all the composite spectrum was shifted into its rest-frame using the mean shift measured for the two chosen photospheric stellar lines and the two nebular carbon emission lines, weighted for their uncertainties (see Section \ref{sec:Uncertainties}), of $\sim 135 \pm 22$ km/s. \footnote[5]{The error in a weighted mean is defined as $\sigma^2 = 1 / \sum(1/\sigma_i^2)$, where $\sigma_i$ are the measurement uncertainties associated with the stellar and nebular lines.}. 


\subsection{Velocity offset measurements}
\label{sec:Velocity offset measurements}

The different kinds of lines observed in the UV rest-frame spectrum of a SFG were already described in a previous section.
Here we recall that, in the $1200~ - 2900~\AA$ range, seven strong low-ionization interstellar absorption lines, tracers of the large-scale winds, are detectable, one of which is a blend; three low-ionization doublets are also detectable (Fe II and Mg II), in addition to three high-ionization ones (Si IV, C IV, and Al III). The high-ionization doublets do not provide robust tools in the search for outflows, since at our spectral resolution an attempt to separate the ISM absorption from other components, such as the stellar wind P-Cygni profile, is impracticable.

Therefore, in our analysis we concentrated only on the low-ionization interstellar absorption lines and doublets; in particular: Si II $\lambda1260~\AA$, C II $\lambda1334~\AA$, Si II $\lambda1526~\AA$, Fe II $\lambda1608~\AA$, Al III $\lambda1670~\AA$, and Fe II $\lambda2344~\AA$, and the doublets Fe II $\lambda2374,2382~\AA$, Fe II $\lambda2586,2600~\AA$, and Mg II $\lambda\lambda2796,2803~\AA$. 
The O + Si II $\lambda1303~\AA$ feature was excluded because it was impossible to correctly determine the position of the centroids of the two blended lines. The other cited interstellar lines were all detectable and strong enough to ensure a quite good determination of their centroid.

Fe II $\lambda1608~\AA$ centroid is probably the least accurate, since it is the weakest of all the IS lines and its profile is broader and more asymmetric, these all being factors that may alter the position of the centroid during the fitting procedure. Moreover, we suspect that a blend of absorption features (Fe III, Al III, N II) typical of O and B stars in the 1600-1630 $~\AA$ range \citep[see][]{1993ApJS...86....5K} could play a major role in distorting the Fe II $\lambda1608~\AA$ line profile.

\subsection{Uncertainties}
\label{sec:Uncertainties}

For each line, two possible error sources related to the centroid measurement were estimated.
The error due to spectral noise was computed by creating an artificial template with a continuum slope similar to that of the real composite spectrum and perfect gaussian spectral lines of similar intensities to those in the composite. Random noise was then added to the template until the S/N of the real composite spectrum was matched, and the centroids of the spectral lines of interest were measured. For each line, the r.m.s. of thirty repeated measurements, with respect to the input value, was then adopted as the measurement uncertainty. For the strongest absorption lines, the order of magnitude of this uncertainty is $\sim 15 - 20$ km/s; weaker lines have a higher measurement uncertainty with a maximum of $\sim 100$ km/s for the O IV $\lambda1343~\AA$.

We then evaluated how slight variations in the number of constituents of the composite spectrum would affect the aspect of the spectral lines or the position of their centroid, by creating about a dozen composite spectra and omitting from the procedure a number of single spectra randomly chosen between 5 and 10. The lines of interest were then measured in all the spectra: the computed r.m.s. of the measurements for each line was eventually defined as the "galaxy sampling uncertainty" and added quadratically to the aforementioned measurement error. On average, the galaxy sampling uncertainty is $\sim 10$ km/s for the interstellar and nebular lines and  $\sim 35$ km/s for the photospheric stellar ones.

Finally, for unresolved C III] $\lambda\lambda1906,1908~\AA$ doublet, as already mentioned, we fixed the intensities ratio of the two components to be 1.5, but we added to the line measurement an error component of $\sim 39$ km/s to take into account the uncertainty in the real value of flux intensity ratio. For example, \citet{quider2009} measured a flux ratio $F(1906)/F(1908) = 1.1 \pm 0.2$ in the spectrum of the "Cosmic Horshoe", a gravitationally lensed star-forming galaxy at $z \sim 2$, whose physical properties such as SFR, mass, and dust extiction are similar to the average ones computed for our sample. We note that using a flux ratio of 1.1 instead of 1.5, would not change our results.

The total uncertainty for each line is quoted in Table~\ref{tabOffAve}.

\section{Results}
\label{sec:Results}

\begin{figure*}
\centering
\includegraphics[scale=0.8]{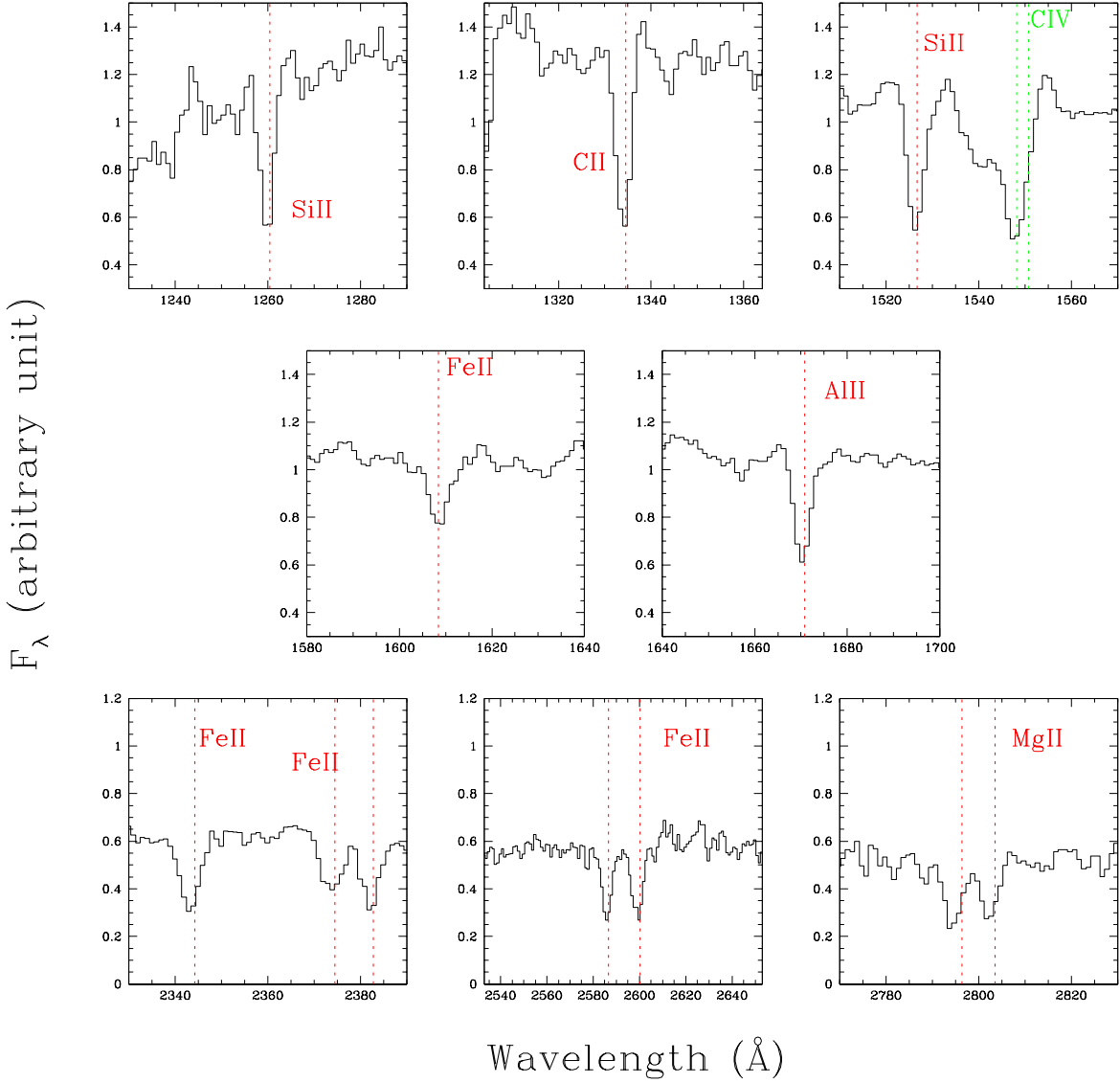}
\caption{Composite spectrum of the 74 spectra of the SFG sample (see Figure~\ref{aveTOTv2_label}): zoom-in of the strongest low-ionization interstellar absorption lines. The C IV high-ionization doublet is also labeled. The spectrum has been shifted into its rest-frame using the velocity correction derived from the stellar/nebular lines ($\sim$ 135 km/s). The vertical lines indicate the expected positions of the line.}
\label{prof_IS}
\end{figure*}
\begin{table*}
\caption[]{Velocity offsets and EWs of strong low-ionization lines, measured in the composite spectrum of the full SFG sample.}
\centering
\begin{tabular}{c c c c c c c}
\hline\hline
Ion 	& Vacuum 	     & IE (eV)\tablefootmark{a} & \multicolumn{2}{c}{EW ($\AA$)\tablefootmark{b}} & \multicolumn{2}{c}{\emph{v} (km/s)}\\
	& wavelength ($\AA$) &         & GMASS & LBG\tablefootmark{c} & GMASS & LBG\tablefootmark{c}\\
\hline\hline          
Si II 	                & 1260.42	& 8.15  & 1.47 $\pm$ 0.09 & 1.63 $\pm$ 0.10 & -79  $\pm$ 19 & -110 $\pm$  60\tablefootmark{e}\\
O I - SiII\tablefootmark{d}& 1302.17, 1304.37  	& -  & 2.43 $\pm$ 0.26 & 2.20 $\pm$ 0.12 & -123 $\pm$ 42 & -270 $\pm$  60\tablefootmark{e}\\   
C II 	                & 1334.53 	& 11.26 & 2.09 $\pm$ 0.15 & 1.72 $\pm$ 0.11 & -105 $\pm$ 33 & -150 $\pm$  60\tablefootmark{e}\\
Si II 	                & 1526.71 	& 8.15  & 1.94 $\pm$ 0.08 & 1.72 $\pm$ 0.18 & -79  $\pm$ 19 & -110 $\pm$  60\tablefootmark{e}\\
Fe II 	                & 1608.45 	& 7.87  & 1.31 $\pm$ 0.11 & 0.91 $\pm$ 0.15 & 24   $\pm$ 40 & -60  $\pm$  60\tablefootmark{e}\\
Al II 	                & 1670.79 	& 5.99  & 1.54 $\pm$ 0.08 & 1.04 $\pm$ 0.15 & -78  $\pm$ 27 & -100 $\pm$  60\tablefootmark{e}\\
Fe II 			& 2344.21 	& 7.87  & 2.45 $\pm$ 0.13 &  & -139 $\pm$ 27 &\\ 
Fe II\tablefootmark{f}	& 2374.46 	& 7.87  & 1.87 $\pm$ 0.13 &  & -72  $\pm$ 32 &\\ 
Fe II\tablefootmark{f}	& 2382.76 	& 7.87  & 2.21 $\pm$ 0.15 &  & -72  $\pm$ 27 &\\ 
Fe II\tablefootmark{g}	& 2586.65 	& 7.87  & 2.65 $\pm$ 0.17 &  & -108 $\pm$ 27 &\\ 
Fe II\tablefootmark{g}	& 2600.17 	& 7.87  & 2.77 $\pm$ 0.14 &  & -108 $\pm$ 28 &\\ 
Mg II\tablefootmark{h}	& 2796.35 	& 7.65  & 2.43 $\pm$ 0.10 &  & -176 $\pm$ 27 &\\ 
Mg II\tablefootmark{h}	& 2803.53 	& 7.65  & 2.07 $\pm$ 0.10 &  & -176 $\pm$ 27 &\\ 
\hline\hline
\end{tabular}
\tablefoot{
\tablefoottext{a}Ionization energy.\\
\tablefoottext{b}Rest-frame equivalent widths.\\
\tablefoottext{c}Values from \citet{2003ApJ...588...65S}.\\
\tablefoottext{d}All the measurements, both ours and Shapley's, refer to the blend of the two lines.\\
\tablefoottext{e}\citet{2003ApJ...588...65S} quotes only the mean error in the velocity offsets.\\
\tablefoottext{f}\tablefoottext{g}\tablefoottext{h}Doublets.\\
}
\label{tabOffAve}
\end{table*}

Figure~\ref{prof_IS} shows zoomed-in spectral regions containing the main low-ionization interstellar absorption lines whose offsets were measured. The spectrum has been shifted into its rest-frame defined by the velocity correction of $\sim$ 135 km/s computed from the stellar and nebular lines.

Most of the lines are blueshifted with respect to their expected positions, as indicated by the vertical dashed lines: this confirms the presence of outflows in the galaxies of our sample, at least on average. These results, qualitatively shown in the figures, are presented in Table~\ref{tabOffAve}. A mean velocity offset of $\sim 100$ km/s was computed for all the IS lines in our composite. The shift velocities measured for the single low-ionization lines in the composite spectrum are listed, along with the results from \citet{2003ApJ...588...65S} for the composite spectrum of a sample of Lyman-break selected galaxies (LBG) at $\emph{z} \sim 3$. The EWs of all the lines were also measured, and presented.
It was impossible to compare offsets and EWs of all the lines, since our composite spectrum extends more to the red than the LBG composite.

In the wavelength range sampled by both spectra, given the uncertainties, the two sets of velocities are fairly consistent, probing that gas outflows are a common phenomenon in SFGs at all redshifts.

On the other hand, the EWs measured in our composite are larger than in the LBG composite. Looking at the colour excess in the two samples, the LBG sample has $E(B-V)$ values that are, on average, $\sim 0.1$ mag lower than the ones computed for our sample. This can be intepreted as evidence of the correlation between the interstellar absorption line strength and dust extinction.

\subsection{Searching for correlations between UV spectroscopic properties and galaxy physical properties}
\label{sec:Searching for correlations between UV spectroscopic properties and galaxy physical properties}

\begin{figure*}
\centering
\includegraphics[scale=0.3]{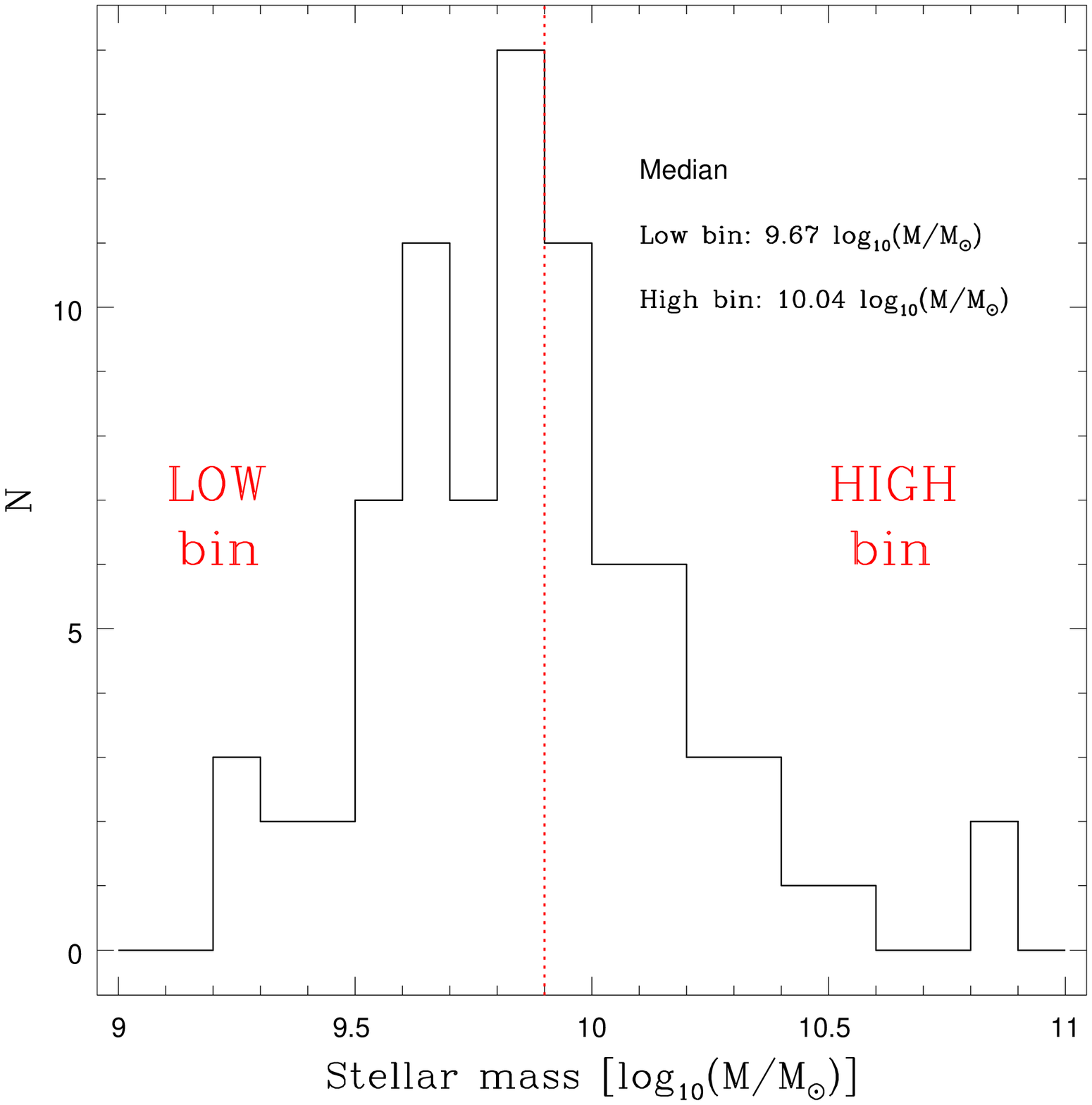}
\includegraphics[scale=0.3]{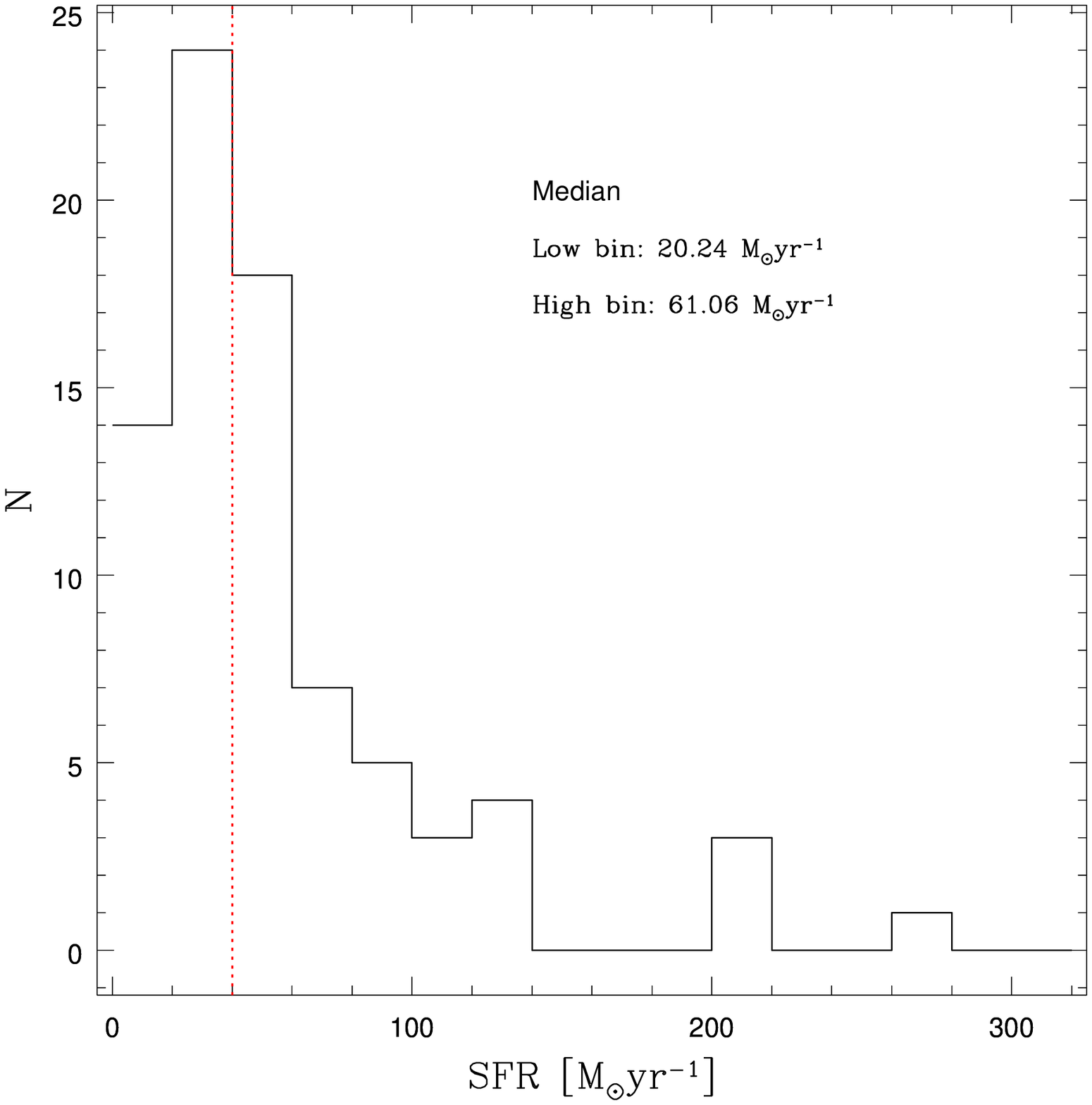}
\includegraphics[scale=0.3]{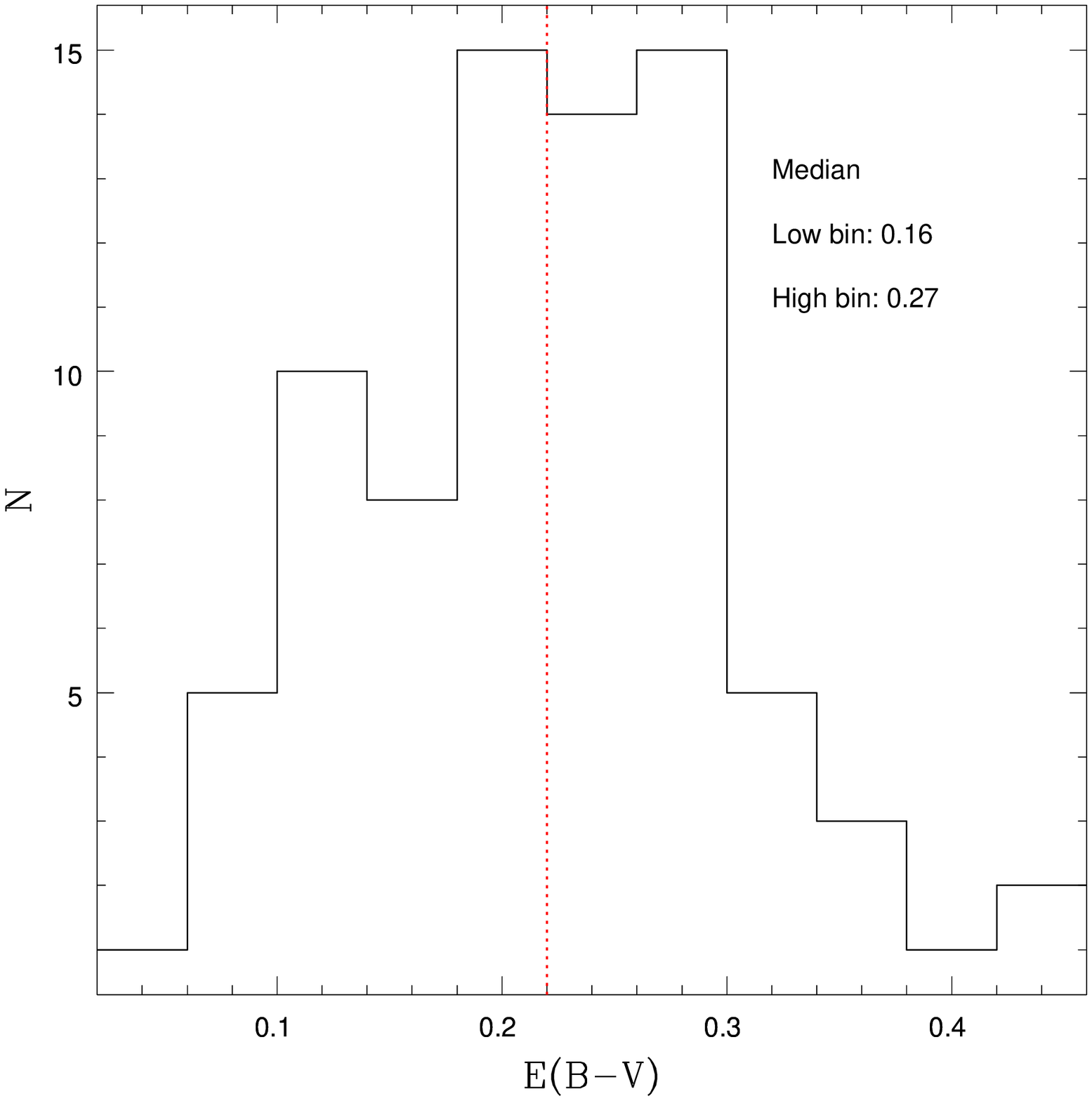}
\caption{Distribution over the sample of stellar mass, SFR (from 1500$~\AA$ luminosity measured on the spectra with dust extinction derived from the continuum slope) and E(B-V) (computed from the UV continuum slope). The median value of both the sub-samples is also reported in each panel.}
\label{hist}
\end{figure*}

\begin{table*}
\caption[]{Equivalent widths and velocity offsets of the strongest interstellar absorption lines (plus the He II $\lambda1640~\AA$ emission line), measured in composite spectra.}
\centering
\begin{tabular}{c c c c c c c c}
\hline\hline
Ion & Vacuum               & EW & \emph{v} & EW & \emph{v} (km/s) & EW$_{ratio}$\tablefootmark{a} & $\Delta$\emph{v}\tablefootmark{b}\\
    & wavelength ($\AA$) & ($\AA$) & (km/s) & ($\AA$) & (km/s) & ($\AA$) & (km/s)\\
\hline\hline		    	    			
\multicolumn{2}{c}{SFR} & \multicolumn{2}{c}{Bin 1\tablefootmark{c}:$~ $  20.24 M$_\odot~$yr$^{-1}$} & \multicolumn{2}{c}{Bin 2\tablefootmark{c}:$~ $  61.06 M$_\odot~$yr$^{-1}$}\\
\hline                						    
Si II & 1260.42 & 0.97 $\pm$ 0.09 & -97  $\pm$ 19 & 1.57 $\pm$ 0.09 & -145 $\pm$ 19 & 1.62 $\pm$ 0.33 & -49 $\pm$ 27 \\ 	  
C II  & 1334.53 & 1.98 $\pm$ 0.15 & -114 $\pm$ 33 & 2.17 $\pm$ 0.15 & -138 $\pm$ 33 & 1.10 $\pm$ 0.33 & -24 $\pm$ 47 \\ 	  
Si II & 1526.71 & 2.44 $\pm$ 0.08 & -97  $\pm$ 19 & 3.73 $\pm$ 0.08 & -145 $\pm$ 19 & 1.53 $\pm$ 0.27 & -49 $\pm$ 27 \\ 	  
Fe II & 1608.45 & 1.09 $\pm$ 0.11 & 9	 $\pm$ 40 & 1.21 $\pm$ 0.11 & -47  $\pm$ 40 & 1.11 $\pm$ 0.25 & -56 $\pm$ 57 \\ 	  
He II & 1640.42 & -1.13 $\pm$ 0.10 &              & -0.87 $\pm$ 0.10 &              & 0.77 $\pm$ 0.16 &              \\ 
Al II & 1670.79 & 1.35 $\pm$ 0.08 & -108 $\pm$ 27 & 1.77 $\pm$ 0.08 & -95  $\pm$ 27 & 1.31 $\pm$ 0.22 & 13  $\pm$ 38 \\ 	  
Fe II & 2344.21 & 2.19 $\pm$ 0.13 & -185 $\pm$ 27 & 3.02 $\pm$ 0.13 & -120 $\pm$ 27 & 1.38 $\pm$ 0.38 & 65  $\pm$ 38 \\ 	  
Fe II & 2374.46 & 1.81 $\pm$ 0.13 & -150 $\pm$ 32 & 2.77 $\pm$ 0.13 & -89  $\pm$ 32 & 1.53 $\pm$ 0.43 & 61  $\pm$ 45 \\ 	  
Fe II & 2382.76 & 2.06 $\pm$ 0.15 & -150 $\pm$ 27 & 2.28 $\pm$ 0.15 & -89  $\pm$ 27 & 1.11 $\pm$ 0.33 & 61  $\pm$ 38 \\ 	  
Fe II & 2586.65 & 1.88 $\pm$ 0.17 & -159 $\pm$ 27 & 3.74 $\pm$ 0.17 & -87  $\pm$ 27 & 1.99 $\pm$ 0.84 & 72  $\pm$ 38 \\ 	  
Fe II & 2600.17 & 2.57 $\pm$ 0.14 & -159 $\pm$ 28 & 3.33 $\pm$ 0.14 & -87  $\pm$ 28 & 1.30 $\pm$ 0.38 & 72  $\pm$ 40 \\ 	  
Mg II & 2796.35 & 2.53 $\pm$ 0.10 & -239 $\pm$ 27 & 2.18 $\pm$ 0.10 & -137 $\pm$ 27 & 0.86 $\pm$ 0.17 & 101 $\pm$ 38 \\ 	  
Mg II & 2803.53 & 1.87 $\pm$ 0.10 & -238 $\pm$ 27 & 2.37 $\pm$ 0.10 & -137 $\pm$ 27 & 1.27 $\pm$ 0.26 & 101 $\pm$ 38 \\ 	  
\hline				    	    		    
\multicolumn{2}{c}{Stellar Mass} & \multicolumn{2}{c}{Bin 1\tablefootmark{c}:$~ $  9.67 log(M$_\odot$)} & \multicolumn{2}{c}{Bin 2\tablefootmark{c}:$~ $  10.04 log(M$_\odot$)}\\		  
\hline  					  
Si II & 1260.42 & 1.61 $\pm$ 0.09 & -140 $\pm$ 19 & 1.89 $\pm$ 0.09 & -138 $\pm$ 19 & 1.17 $\pm$ 0.21 & 2   $\pm$ 27 \\ 	  
C II  & 1334.53 & 2.25 $\pm$ 0.15 & -134 $\pm$ 33 & 2.34 $\pm$ 0.15 & -91  $\pm$ 33 & 1.04 $\pm$ 0.31 & 43  $\pm$ 47 \\ 	  
Si II & 1526.71 & 2.47 $\pm$ 0.08 & -140 $\pm$ 19 & 3.14 $\pm$ 0.08 & -138 $\pm$ 19 & 1.27 $\pm$ 0.21 & 2   $\pm$ 27 \\ 	  
Fe II & 1608.45 & 1.17 $\pm$ 0.11 & 46	 $\pm$ 40 & 1.28 $\pm$ 0.11 & -13  $\pm$ 40 & 1.09 $\pm$ 0.24 & -59 $\pm$ 57 \\ 	  
He II & 1640.42 & -1.23 $\pm$ 0.10 &              & -0.81 $\pm$ 0.10 &              & 0.66 $\pm$ 0.14 &              \\ 
Al II & 1670.79 & 1.33 $\pm$ 0.08 & -80  $\pm$ 27 & 1.87 $\pm$ 0.08 & -92  $\pm$ 27 & 1.41 $\pm$ 0.24 & -12 $\pm$ 38 \\ 	  
Fe II & 2344.21 & 2.37 $\pm$ 0.13 & -149 $\pm$ 27 & 2.72 $\pm$ 0.13 & -139 $\pm$ 27 & 1.15 $\pm$ 0.30 & 9   $\pm$ 38 \\ 	  
Fe II & 2374.46 & 1.68 $\pm$ 0.13 & -82  $\pm$ 32 & 2.51 $\pm$ 0.13 & -93  $\pm$ 32 & 1.49 $\pm$ 0.42 & -11 $\pm$ 45 \\ 	  
Fe II & 2382.76 & 1.71 $\pm$ 0.15 & -82  $\pm$ 27 & 2.90 $\pm$ 0.15 & -93  $\pm$ 27 & 1.7  $\pm$ 0.58 & -11 $\pm$ 38 \\ 	  
Fe II & 2586.65 & 2.59 $\pm$ 0.17 & -139 $\pm$ 27 & 3.31 $\pm$ 0.17 & -101 $\pm$ 27 & 1.28 $\pm$ 0.45 & 37  $\pm$ 38 \\ 	  
Fe II & 2600.17 & 2.88 $\pm$ 0.14 & -139 $\pm$ 28 & 3.27 $\pm$ 0.14 & -101 $\pm$ 28 & 1.14 $\pm$ 0.32 & 37  $\pm$ 40 \\ 	  
Mg II & 2796.35 & 2.24 $\pm$ 0.10 & -183 $\pm$ 27 & 2.37 $\pm$ 0.10 & -209 $\pm$ 27 & 1.06 $\pm$ 0.21 & -26 $\pm$ 38 \\ 	  
Mg II & 2803.53 & 1.63 $\pm$ 0.10 & -183 $\pm$ 27 & 2.66 $\pm$ 0.10 & -209 $\pm$ 27 & 1.63 $\pm$ 0.37 & -26 $\pm$ 38 \\ 
\hline				    	    		    
\multicolumn{2}{c}{E(B-V)} & \multicolumn{2}{c}{Bin 1\tablefootmark{c}:$~ $  0.16} & \multicolumn{2}{c}{Bin 2\tablefootmark{c}:$~ $  0.27}\\		  
\hline  					  
Si II & 1260.42 & 1.43 $\pm$ 0.09 & -84  $\pm$ 19 & 2.25 $\pm$ 0.09 & -38  $\pm$ 19 & 1.57 $\pm$ 0.31 & 46  $\pm$ 27 \\ 	  
C II  & 1334.53 & 1.85 $\pm$ 0.15 & -160 $\pm$ 33 & 2.18 $\pm$ 0.15 & -58  $\pm$ 33 & 1.18 $\pm$ 0.36 & 102 $\pm$ 47 \\ 	  
Si II & 1526.71 & 1.70 $\pm$ 0.08 & -84  $\pm$ 19 & 2.27 $\pm$ 0.08 & -38  $\pm$ 19 & 1.34 $\pm$ 0.22 & 46  $\pm$ 27 \\ 	  
Fe II & 1608.45 & 1.25 $\pm$ 0.11 & 7	 $\pm$ 40 & 1.61 $\pm$ 0.11 & 86   $\pm$ 40 & 1.29 $\pm$ 0.29 & 78  $\pm$ 57 \\ 	  
He II & 1640.42 & -0.88 $\pm$ 0.10 &              & -0.96 $\pm$ 0.10 &              & 1.10 $\pm$ 0.22 &              \\ 
Al II & 1670.79 & 1.31 $\pm$ 0.08 & -46  $\pm$ 27 & 1.81 $\pm$ 0.08 & -51  $\pm$ 27 & 1.38 $\pm$ 0.23 & -5  $\pm$ 38 \\ 	  
Fe II & 2344.21 & 2.28 $\pm$ 0.13 & -103 $\pm$ 27 & 2.76 $\pm$ 0.13 & -110 $\pm$ 27 & 1.21 $\pm$ 0.32 & -7  $\pm$ 38 \\ 	  
Fe II & 2374.46 & 1.59 $\pm$ 0.13 & -78  $\pm$ 32 & 2.15 $\pm$ 0.13 & -21  $\pm$ 32 & 1.35 $\pm$ 0.37 & 57  $\pm$ 45 \\ 	  
Fe II & 2382.76 & 1.96 $\pm$ 0.15 & -78  $\pm$ 27 & 2.40 $\pm$ 0.15 & -21  $\pm$ 27 & 1.23 $\pm$ 0.38 & 57  $\pm$ 38 \\ 	  
Fe II & 2586.65 & 1.53 $\pm$ 0.17 & -98  $\pm$ 27 & 3.89 $\pm$ 0.17 & -58  $\pm$ 27 & 2.55 $\pm$ 1.28 & 39  $\pm$ 38 \\ 	  
Fe II & 2600.17 & 2.11 $\pm$ 0.14 & -98  $\pm$ 28 & 3.58 $\pm$ 0.14 & -58  $\pm$ 28 & 1.70 $\pm$ 0.54 & 39  $\pm$ 40 \\ 	  
Mg II & 2796.35 & 1.58 $\pm$ 0.10 & -186 $\pm$ 27 & 3.04 $\pm$ 0.10 & -124 $\pm$ 27 & 1.93 $\pm$ 0.47 & 63  $\pm$ 38 \\ 	  
Mg II & 2803.53 & 1.64 $\pm$ 0.10 & -186 $\pm$ 27 & 2.32 $\pm$ 0.10 & -123 $\pm$ 27 & 1.41 $\pm$ 0.30 & 63  $\pm$ 38 \\ 	  
\hline\hline

\end{tabular}
\tablefoot{
\tablefoottext{a}EW$_{ratio}$ = EW$_{High Bin}$ / EW$_{Low Bin}$\\
\tablefoottext{b}$\Delta$\emph{v} = \emph{v}$_{High Bin}$ - \emph{v}$_{Low Bin}$\\
\tablefoottext{c}Median of each bin.\\
}
\label{tabVelOff}
\end{table*}
\begin{table*}
\caption[]{Mean measured equivalent widths and velocity offsets.}
\centering
\begin{tabular}{l c c c c c c}
\hline\hline
    				& \multicolumn{2}{c}{EW$_{IS}$($\AA$)}  		& \multicolumn{2}{c}{\emph{v$_{IS}$}(km/s)} & EW$_{ratio}$\tablefootmark{b} & $\Delta$\emph{v}\tablefootmark{c}\\
\hline\hline               				  		
SFR (M$_\odot$~yr$^{-1}$)		& Bin 1\tablefootmark{a}:$~ $ 20.24 		& Bin 2\tablefootmark{a}:$~ $ 61.06 		& Bin 1\tablefootmark{a}:$~ $ 20.24		& Bin 2\tablefootmark{a}:$~ $ 61.06 && \\
						& 1.90 $\pm$ 0.15	& 2.51 $\pm$ 0.24	& -141 $\pm$ 19		& -110 $\pm$ 9 & 1.33 $\pm$ 0.16 & 31 $\pm$ 21\\
\hline                						    
Stellar Mass (log(M$_\odot$))	& Bin 1\tablefootmark{a}:$~ $ 9.67 			& Bin 2\tablefootmark{a}:$~ $ 10.04\ 		& Bin 1\tablefootmark{a}:$~ $ 9.67 			& Bin 2\tablefootmark{a}:$~ $ 10.04 && \\  
						& 1.99 $\pm$ 0.16	& 2.52 $\pm$ 0.18	& -117 $\pm$ 18		& -118 $\pm$ 16 & 1.26 $\pm$ 0.13 & -1 $\pm$ 24\\
\hline                						    
E(B-V)	& Bin 1\tablefootmark{a}:$~ $ 0.16 			& Bin 2\tablefootmark{a}:$~ $ 0.27 		& Bin 1\tablefootmark{a}:$~ $ 0.16 			& Bin 2\tablefootmark{a}:$~ $ 0.27 && \\  
						& 1.69 $\pm$ 0.09	& 2.52 $\pm$ 0.20	& -99 $\pm$ 16		& -51 $\pm$ 16 & 1.50 $\pm$ 0.14 & 48 $\pm$ 23\\
\hline\hline
\end{tabular}
\tablefoot{
\tablefoottext{a}Median of each bin.\\
\tablefoottext{b}EW$_{ratio}$ = EW$_{High Bin}$ / EW$_{Low Bin}$\\
\tablefoottext{c}$\Delta$\emph{v} = \emph{v}$_{High Bin}$ - \emph{v}$_{Low Bin}$\\
}
\label{tabAveVelOff}
\end{table*}

The spectral analysis of the composite spectrum has provided average information about our entire sample of $z \sim 2$ high-quality star-forming galaxies. The next step was to investigate possible correlations between galaxy UV spectral characteristics (EW, outflows) and galaxy general properties, in particular stellar mass, SFR, and dust-extinction.

Stellar masses were computed by means of the SED fitting of \citet{2005MNRAS.362..799M} models, while the colour excess was derived from the continuum slope and the SFR was obtained from UV spectroscopic luminosities, as described in Section \ref{sec:Star-formation rate}.

Our sample was divided into two equally populated bins, for each property, and a composite spectrum of each group of galaxies was created, following the same procedure adopted in the creation of the composite of the total sample. 
We decided to work with only two bins because of the relatively small number of spectra at our disposal. As already stated, to have reliable measurements of centroids and EWs of the spectral lines of interest, a high S/N is needed in the composite spectra, and the composition procedure is the more effective in improving the S/N the higher the number of single spectra involved. 

Figure~\ref{hist} shows the distribution of the chosen physical properties over our sample, with the binning division values marked and the median values of each bin reported. The composite spectrum of each bin was shifted into the rest-frame given by the stellar and nebular lines. Then, the velocity offsets of the low-ionization absorption lines were measured, which are presented in Table~\ref{tabVelOff}. Since Table~\ref{tabVelOff} is a bit "dense", the mean velocity offset and EW of the interstellar low-ionization absorption lines in each bin are reported in Table~\ref{tabAveVelOff}.

\subsubsection{Outflow velocities}

The IS lines show a significant velocity shift in all the composite spectra ($\delta v_{IS} \sim 100$ km/s), but there seem to be no evident correlations between the velocity offsets of the interstellar absorption lines and the physical properties of the galaxies in our sample, apart from some weak differences in the colour excess and SFR.

However, our results do not completely exclude the existence of these correlations, since we are aware of some factors that may hide them. 
First of all, the magnitude of the uncertainties might be playing a significant role in hiding real weak correlations. Moreover, the distributions of SFR, stellar mass, and colour excess do not favor a division of the sample into two equally populated bins, since most of the galaxies are concentrated close to the threshold value, and the errors are large enough to cause confusion between the two bins. Finally, the fact has to be considered that there may be absorption also from gas which is not outflowing, expecially in more massive galaxies \citep{steidel2010}. This $v \sim 0$ component may affect the emergent line profile in the sense that the velocities computed from the centroid shift of the IS lines may be underestimated.

\subsubsection{Equivalent widths of outflow-related absorption lines}

The EW measurements show significant differences, when comparing the composite spectra of different sub-samples. A positive trend between the EWs of the interstellar absorption lines and SFR, stellar mass, and E(B-V) is suggested, which is the main result of this last part of our analysis.

In other samples of star-forming galaxies, at different redshifts but with comparable properties of stellar mass and SFR, interstellar absorption lines associated with large scale outflows have been found to be stronger in galaxies with higher SFR, higher stellar mass, and more heavily extincted \citep{2003ApJ...588...65S, 2009ApJ...692..187W, 2010ApJ...719.1503R}. Given the correlations between these galaxy properties, it is not possible to establish the characteristic with which inter-stellar lines parameters are more strongly connected, even if the most significant of the three correlations appears to be the one with E(B-V).

From these results it may be inferred that the EWs of the absorption inter-stellar lines are probably related to the velocity dispersion of the gas, or to its geometry, or to a combination of the two effects, all of which are difficult to distinguish with the actual data.

Finally, we report the EW of He II $\lambda1640~\AA$ to be stronger in the composite spectra of the less massive and star-forming galaxies.

\section{Summary and conclusions}

We have studied a sample of 74 star-forming galaxies at $z \sim 2$ from GMASS by analyzing their rest-frame UV spectra. The first part of this work focused on the determination of dust extinction and SFR for the galaxies in our sample, while in the second part large-scale gas-outflow-related properties were also analyzed. 
The main results of these analyses may be summarized as follows. 
\begin{enumerate}
      \item The continuum slope of the 74 individual spectra of our sample was measured, following \citet{1994ApJ...429..582C}, with a mean value for the sample of $<\beta>=-1.11 \pm 0.44$ (r.m.s.); this value confirms the results of similar studies in the same redshift range of our sample, which are also consistent with the findings in the local Universe.
      \item Dust extinction was derived for each galaxy from the continuum slope of its spectrum, and used to correct the flux measured at 1500 $\AA$ (rest-frame). The corrected flux was then converted into SFR by applying the relation from \citet{1998ARA&A..36..189K} scaled for a Kroupa IMF; the galaxies of our sample show average $<SFR>~=52 \pm 48~M_\odot~ yr^{-1}$ (r.m.s.).
      \item A positive correlation between SFR and stellar mass was found, with a logarithmic slope of the relation of $1.10 \pm 0.10$, in agreement with previous studies; the SSFR was also found to decrease with stellar mass. 

      \item A composite spectrum was created by averaging the high-quality spectra of 74 SFGs, to obtain average information about large-scale gas-outflow-related properties. The low-ionization absorption lines associated with the interstellar medium were found to be blueshifted, with respect to the rest frame of the system, indicating the presence of outflowing gas with velocities of the order of $\sim 100$ km/s. 
Our measured velocity offsets were found to be fairly consistent with that obtained by \citet{2003ApJ...588...65S} for the composite spectrum of a sample of LBGs at $z\sim 3$, in the wavelength range sampled by both spectra (between 1000 and 2000 $\AA$).
      \item The EWs of the low-ionization absorption lines associated with the interstellar medium were also measured: the EWs measured in our composite appear to be larger than in the LBG composite. This could be interpreted as evidence of a trend between the interstellar absorption-line strength and dust extinction, since the LBGs have $E(B-V)$ values that are, on average, $\sim 0.1$ mag lower than the ones computed for our sample.
      \item In the composite spectrum of the galaxies showing $Ly\alpha$ in emission, a velocity difference of $\sim 566 \pm 27$km/s between this line and the IS absorption ones was measured. Moreover, the comparison of composite spectra of two subsamples, one with $Ly\alpha$ in emission, the other with the feature in absorption, showed that the continuum slope is bluer in the first one (respectively, $\beta_{emLy\alpha} = -1.40 \pm 0.06$ and $\beta_{absLy\alpha} = -1.04 \pm 0.03$), thus confirming the existence of a relation between UV continuum slope, and therefore dust extinction, and the $Ly\alpha$ profile in SFGs.
      \item Possible correlations between galaxy UV spectral characteristics (EW, outflows) and galaxy general properties, in particular stellar mass, SFR, and dust extinction, were investigated by dividing the sample into bins, for each of the aforementioned properties. No evidence of significant correlation was found between the velocity offsets of the interstellar absorption lines and the physical properties of the galaxies in our sample. However, this does not completely exclude the existence of such correlations, since the magnitude of the uncertainties may hide them.
      \item The He II $\lambda1640~\AA$ emission line, which is a feature of winds from WR stars, was measured in all the composite spectra and the values were compared with the predictions of the most recent stellar population synthesis models. Given the physical properties of the galaxies of our sample, the models predict EWs which are fairly consistent with our measurements. We also report the He II $\lambda1640~\AA$ to be stronger in the composite spectra of the less massive and star-forming galaxies, a puzzling result since the higher fraction of WR stars is expected at higher metallicities. 
      \item A positive trend between the EWs of the interstellar absorption lines and SFR, stellar mass, and E(B-V) has been found, which has been reported to also hold at different redshifts. The most significant of the three correlations appears to be the one with E(B-V), which suggests that the EWs of the absorption inter-stellar lines are probably related to the velocity dispersion of the gas, to its geometry, or to a combination of the two effects.
\end{enumerate}

New data provided by Herschel will allow deeper investigations of the SFR in the high-redshift galaxies of the GMASS sample and, consequently, help us to constrain the evolution of the star formation over cosmic time.

Moreover, to improve our understanding of the phenomenon of large-scale outflows, deeper spectroscopic surveys will be needed in the future. In particular, it would be extremely useful to have multi-wavelength spectral data. High-quality spectra sampling not only the rest-frame UV range, but also longer wavelengths, could allow a robust determination of systemic velocities and, consequently, the study of the warm phase of large-scale outflows in single galaxies, a task that to date has only been attempted for lens-magnified objects. 



\begin{acknowledgements}
The authors wish to thank the anonymous referee for his/her comments and suggestions.
MT thanks P.A.R. members for useful discussions.
JK is supported by the German Science Foundation (DFG) via German-Israeli Project Cooperation grant STE1869/1-1.GE625/15-1.
\end{acknowledgements}


\bibliographystyle{aa} 
\bibliography{references} 

%

\end{document}